\documentclass[pre,aps,showpacs,twocolumn,floatfix,superscriptaddress]{revtex4}

\usepackage{graphicx}
\usepackage{bm} 
\usepackage{epsfig}
\usepackage{color}
\usepackage{enumitem}
\usepackage[hidelinks]{hyperref}
\usepackage{amsmath}
\usepackage{float}%
\usepackage{hyperref}
\usepackage{siunitx}
\usepackage{booktabs}

\setlist[itemize]{leftmargin=*}
\setlist[enumerate]{leftmargin=*}	
\newcommand{\be}{\begin{equation}} 
\newcommand{\ee}{\end{equation}}
\newcommand{\bea}{\begin{eqnarray}}
\newcommand{\eea}{\end{eqnarray}}
\newcommand{\ba}{\begin{array}}
\newcommand{\ea}{\end{array}}

\begin{document}

\title{Polymer translocation driven by longitudinal and transversal time-dependent end-pulling forces.}

\author{A. S\'ainz-Agost}
\email{asainz@unizar.es}
\affiliation{Dpto. de F\'isica de la Materia Condensada,
Universidad de Zaragoza. 50009 Zaragoza, Spain}
\affiliation{Instituto de Biocomputaci\'on y F\'isica de Sistemas
Complejos, Universidad de Zaragoza. 50018 Zaragoza, Spain}

\author{F. Falo}
\email{fff@unizar.es}
\affiliation{Dpto. de F\'isica de la Materia Condensada,
Universidad de Zaragoza. 50009 Zaragoza, Spain}
\affiliation{Instituto de Biocomputaci\'on y F\'isica de Sistemas
Complejos, Universidad de Zaragoza. 50018 Zaragoza, Spain}

\author{A. Fiasconaro}
\email{afiascon@unizar.es}
\affiliation{Dpto. de F\'isica de la Materia Condensada,
Universidad de Zaragoza. 50009 Zaragoza, Spain}
\affiliation{Instituto de Biocomputaci\'on y F\'isica de Sistemas
Complejos, Universidad de Zaragoza. 50018 Zaragoza, Spain}
\affiliation{Istituto di Biofisica, Consiglio Nazionale delle Ricerche. Palermo, Italy}

\date{\today}

\begin{abstract}
Polymer translocation has long been a topic of interest in the field of biological physics given its relevance in both biological (protein and DNA/RNA translocation through nuclear and cell membranes) and technological processes (nanopore DNA sequencing, drug delivery). 

In this work, we simulate the translocation of a semiflexible homopolymer through an extended pore, driven by both a constant and a time-dependent end-pulled force, employing a model introduced in previous studies. The time dependence is simplistically modeled as a cosine function, and we distinguish between two scenarios for the driving --longitudinal force and transversal force-- depending on the relative orientation of the force, parallel or perpendicular respectively, with respect to the pore axis. Beside some key differences between the two drivings, the mean translocation times present a large minimum region as function of the frequency of the force that is typical of the Resonant Activation effect. The presence of the minimum is independent on the elastic characteristics of the polymeric chains and reveals a linear relation between the optimum mean translocation time and the corresponding period of the driving. The mean translocation times show different scaling exponent with the polymer length for different flexibilities. Lastly, we derive an analytical expression of the mean  translocation time for low driving frequency, which clearly agrees with the simulations.
\end{abstract}


\maketitle

\section{Introduction.}
Polymer translocation has been a subject of significant interest in both soft-matter and biological physics due to its relevance in many biological and technological processes. For instance, transportation of DNA, RNA, and proteins through cell and nuclear membranes \cite{1996_kasianowicz_PNAS_characterization_nucleotides}, protein degradation \cite{2001_Ishikawa_PNAS_clpap}, and DNA virus injection \cite{2007_Lebedev_EMBO_virus_translocation} are examples of the former, while the study of these processes and subsequent fabrication of technological devices at the nanoscale used for a variety of applications, namely DNA sequencing through either biological or artificial nanopores,  \cite{2012_Schneider_nature_sequencing} or biosensing \cite{2018_Lee_AV_nanopore_review, 2013_Stoloff_COB_nanopore_review} exemplify the latter.

The growing interest in this field has given rise to a series of computational models aiming to describe the behavior of translocating bodies under different conditions, with the goal to find new properties and applications for experimental uses and artificial devices. 

The stochastic models of translocation dynamics have increased in complexity over time, ranging from the application of a constant force to the translocating molecule \cite{2019_Mohammadreza_constant_force, 2023_Lu_review_force_current}, where the pore is a passive channel \cite{2010_Schneider_passive_pore} that permits the passage of the molecule  between the two sides of a membrane. The force is here generated by a particle current \cite{2023_Lu_review_force_current} or by an electric field \cite{2018_Hsiao_electric_field}.

A simple unique --eventually time-dependent-- energy barrier has been also used to describe the overall translocation~\cite{2010_Fiasconaro_Spagnolo_PB_resonant_activation} or a ratchet-like potential to describe the driving mechanism has also been implemented~\cite{2013_Fiasconaro_Falo_PRE_phi29}.
In the last decade, active pores, {\it i.e.} time-dependent pore drivings, have been introduced~\cite{2010_Fiasconaro_Falo_PRE_mid_point, 2011_cohen_flickering_pores,2011_Rowghanian_translocation_constant_force_pore, 2013_Fiasconaro_Falo_PRE_phi29, 2015_Sarabadani_time_depdendent_translocation, 2015_fiasconaro_falo_polymer_3D}, up to the more recent edge-pulled polymer that mimics the force spectroscopy experimental setups \cite{2022_Paun_polymer_constant_force, 2008_Neuman_force_spectroscopy}. 

According to their dimensions we distinguish punctual \cite{2009_Sun_point_pore} or spatially extended \cite{2010_Fiasconaro_Falo_PRE_mid_point} pores, the latter being more popular in recent years.

Regarding the functional form of the active driving itself, different options have been considered: sinusoidal forces \cite{2015_fiasconaro_falo_polymer_3D, 2010_Fiasconaro_Falo_PRE_mid_point}, stochastic random telegraph noises (RTN) \cite{2011_Fiasconaro_Falo_JSM_end_pulled_dichotomic} or dichotomous ATP-based noises motivated by the action of molecular motors \cite{2012_Fiasconaro_Falo_MM_ATP,2013_Fiasconaro_Falo_PRE_phi29,2017_Fiasconaro_SR_ATP_motor}. 

\begin{figure}[H]
 \centering
 \includegraphics[width=1.0\linewidth]{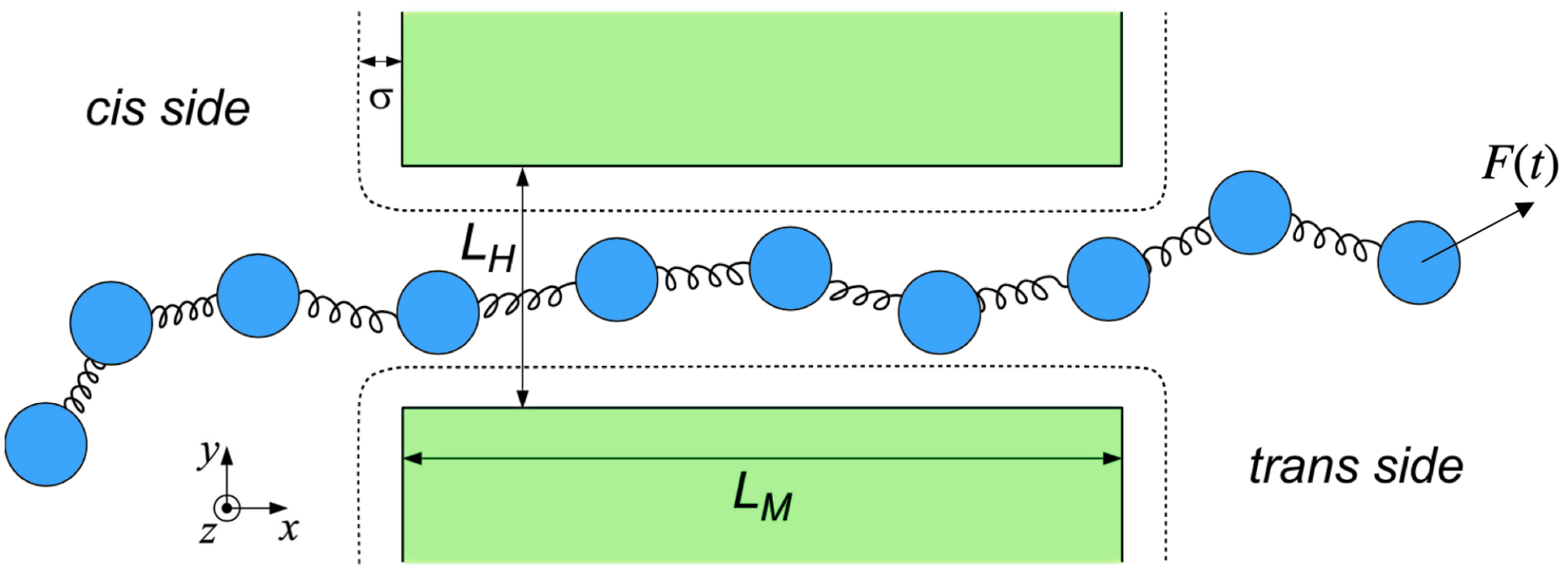}
 \caption{Section of the polymer translocating through a nanopore in the 3d space. The pore has a square section of width $L_H$ and its length is $L_M$, with the same repulsive walls as the whole membrane. The polymer is pulled through the pore with a time dependent force $F(t)$.}
 \label{f-channel}
\end{figure}

Generally speaking, the driving force can be applied to the polymer through two main methods: in the space inside the pore or onto one of its edges. In the first case the driving applies to the monomers moving along the channel~\cite{2010_Fiasconaro_Falo_PRE_mid_point, 2013_Bhattacharya_driven_pore_force_2D, 2012_Saito_pore_force}, either as a result of changes in pressure over the two sides of a membrane, or as the effect of a molecular machine capable of producing movement in the pore. In the second, the force is applied to the first monomer of the chain, the pore thus taking a more passive stand. This end-pulled force \cite{2018_Sarabadani_end_pulled} emulates the action of experimental setups like AFM \cite{2006_Ritort_SME_translocation}, optical or magnetic tweezers \cite{2011_Fiasconaro_Falo_JSM_end_pulled_dichotomic, 2006_Keyser_solid_state_nanopore}.

It is interesting to note that many papers have been studying the translocation process from an analytical point of view. In this context, various results on scaling behavior \cite{1996_Sung_free_energy_analytical, 2011_Rowghanian_translocation_constant_force_pore, 2011_Saito_analytical_scaling, 2012_dubbeldam_forced_scaling}, free energy calculation \cite{1999_muthukumar_polymer_translocation, 2009_Sun_Free_energy, 2016_muthukumar_polymer}, or force propagation \cite{2016_Sakaue_escalado_pubmed, 2018_Menais_PRE_end_pulled, 2022_Sarabadani_tension_propagation} can be found in the literature. 

In this work, we study numerically the application of a {\it time-dependent end-pulled force} to polymer chains of different lengths and flexibilities using a 3D model previously employed~\cite{2015_fiasconaro_falo_polymer_3D,2017_Fiasconaro_SR_ATP_motor}. 

The end-pulled mechanism can be useful to explore and check the mathematical aspects of the single barrier translocation~\cite{2018_fiasconaro_falo_force_spectroscopy}, as well as for the detailed analysis of monomers stepping into the pore, especially for sequencing analysis purposes \cite{2022_Fiasconaro_Falo_Polymer_pore_explicit, 2023_singh_chauhan_sequencing_review}.

Our focus here is to investigate the effects of the periodic driving on the translocation times. The time dependence is simplistically modeled as a cosine function, and we distinguish between two scenarios --longitudinal force and transversal force-- depending on the relative orientation of the force, parallel and perpendicular, with respect to the pore axis. We find some possibilities of optimization in the translocation time already found in previous works \cite{2015_fiasconaro_falo_polymer_3D, 2012_Ikonen_RA}, albeit with remarkable differences due to the new geometry here adopted.

The article is organized as follows: In section \ref{methods} we provide an explanation of the model, along with the associated equations of motion and the units of the experiments. Sections \ref{results_trans} and \ref{results_long} contain the results for the two scenarios considered. The final conclusions are contained in section \ref{conclusions}, while the Appendix (section \ref{appendix}) presents the analytical derivation of the mean translocation times at low frequency for both drivings.

\section{Methods} \label{methods}
\subsection{Polymer model.}
The model depicts a three-dimensional chain formed by $N$ identical monomers joint together by elastic springs, that also interact through bending and excluded-volume potentials. The latter applies both to the rest of the chain and the walls of the pore, as shown in Fig.~\ref{f-channel}. The harmonic potential associated to the elastic interaction is given by
\begin{equation}
	V_{\rm el}(d_{i}) = \frac{k_{\rm el}}{2}\sum_{i = 1}^{N} (l_{i} - l_{0})^2,
\end{equation}
where $k_{\rm el}$ is the elastic constant of the interaction, $\bf{r}_i$ the position vector of the \textit{i-}th monomer, $l_{i} = |\bf{r}_{i+1} - \bf{r}_i |$ the distance between consecutive beads and $l_0$ the equilibrium distance. The bending energy of the chain is described by
\begin{equation}
	V_{\rm b}(\theta_{i}) = -k_{\rm{b}}\sum_{i = 1}^{N} \cos (\theta_{i} - \theta_{0}),
\end{equation}
where $k_{\rm{b}}$ is the bending constant, $\theta_{i}$ represents the angle formed by the three adjacent monomers $i-1$, $i$, and $i+1$, and $\theta_0$ is the equilibrium angle. In this particular model, the equilibrium angle is fixed at 0.

Lastly, the excluded volume effect is encoded via a repulsive Lennard-Jones interaction
\begin{equation}
	V_{\rm LJ} =  4\epsilon \sum_{i\neq j = 1}^{N} \left[ \left(\frac{\sigma}{r_{ij}}\right)^{12} - \left(\frac{\sigma}{r_{ij}}\right)^{6} \right] + \epsilon,
\end{equation}
which applies only when the distance between monomers $i$ and $j$, denoted by $r_{ij}$, is less than or equal to $2^{1/6}\sigma$. For larger distances, the interaction is assumed to be zero. 

With the potentials described previously, the dynamics of each monomer can be described using the overdamped Langevin equation of motion:
 \bea
	m \gamma \dot{\bf{r}}_i = - \nabla_{i}V_{\rm el}(d_i) - \nabla_{i} V_{\rm b}(\theta_{i}) -  \nabla_{i} V_{\rm LJ} +  \nonumber \\
	{\bf{F}}_{\rm pull}(t) + {\bf F}_{\rm wall} +\sqrt{2m\gamma k_{B}T} \xi_{i}(t),
\eea
where $\gamma$ and $m$ are the damping and mass of each monomer respectively, and $\xi_{i}(t)$ represents the white noise associated with thermal fluctuations, verifying that
\bea
\nonumber \langle \xi_{i, \alpha} (t) \rangle = 0 \quad ; \quad \langle \xi_{i, \alpha}(t) \xi_{j, \beta}(t') \rangle = \delta_{i,j} \delta_{\alpha, \beta} \delta_{t, t'}
\eea
with $ i = 1, \cdots, N$ and $\alpha, \beta = x, y, z$.\\

There are two additional terms to mention. The first, ${\bf F}_{\rm wall}$, represents the truncated Lennard-Jones interaction between each monomer and the topology of the porous channel as well as the interaction with the limiting membrane, acting as an excluded volume effect. This is depicted in Fig. \ref{f-channel}. The second term, ${\bf F}_{\rm pull}(t)$, is the time-dependent force applied to the first bead of the chain to induce translocation. It has two components: a constant force applied along the length of the channel and a time-dependent force.
\bea
    \label{driving_f}
	{\bf F}_{\rm pull}(t) &=& F_{\rm const} \hat{x} + F_{\rm var}(t) \hat{r} \nonumber\\
 	F_{\rm var}(t) &=& A \cos(2 \pi \nu t - \psi_{0}),
    \label{Sinusoidal force}
\eea
where $A$ is the amplitude of the periodic force, set at 1.5 in program units (unless otherwise specified in later paragraphs), $\nu$ is the frequency, and $\psi_{0}$ the initial phase, sampled randomly with uniform distribution in the range $[0, 2\pi)$.

The amplitude of the constant component $F_{\rm const}$ is selected to ensure successful polymer translocation in the large majority of experiments even in the absence of the time-dependent driving, while also being low enough not to hidden the effect of the variable component, and is specified later in Sections \ref{results_trans} and \ref{results_long}. As for the time-dependent force itself, as commented before, two scenarios have been studied: transversal driving, which is applied along the $y$ direction ($\hat{r}=\hat{y}$), and longitudinal driving, which is directed along the $x$ axis ($\hat{r}=\hat{x}$).

\subsection{Parameters and units of the model.}

The model has several fixed parameters: the equilibrium distance between adjacent monomers was set $l_0 = 1$, the elastic constant is taken as $k_{\rm el} = 1600$, high enough to represent a non-extending chain, the temperature was taken as $k_{\rm B}T = 0.1$, and for the Lennard-Jones potential $\epsilon = 0.3$ and $\sigma = 0.8$ were chosen. Regarding the amplitudes of the forces applied to the first monomer, the constant term is 0.3 and 1.8 for the transversal and longitudinal cases, respectively, while the time-dependent amplitude is 1.5. Regarding the pore itself, the dimensions were set to comfortably fit approximately 5 monomers inside and keep them relatively straight, with $L_M = 5.5$, and $L_H = 2$, although other options for the dimensions can be found in the literature \cite{2012_Ikonen_RA, 2014_suhonen_PRE_polymer_translocation}. The wall is characterized by the same excluded volume parameters taken for the chain, $\epsilon = 0.3$ and $\sigma = 0.8$.

With these parameters and values in mind, $\epsilon$, $l_{0}$ and $m$ can be chosen as the units of energy, length and mass respectively. This gives the Lennard-Jones timescale as $ t_{\rm LJ} = (ml_0^2/\epsilon)^{(1/2)}$. This scale however has to be compensated due to the overdamped conditions chosen for our system, finding that our time unit is $t_{\rm OD} = \gamma t_{\rm LJ}^2 $.

We can try to establish a relation between our model units and experimental setups. Let us take a DNA chain at room temperature, fixing $k_{\rm B} T = 4.1\,{\rm pN\cdot nm}$. Since $k_{\rm B} T = 0.1$ in our program, the energy unit $\epsilon_0 = 41\,{\rm pN\cdot nm}$. Additionally, we can set $l_{0}=1.875\,{\rm nm}$  and $m=936\,{\rm amu}$~\cite{2012_Ikonen_RA}.

With this, we already have the units of time as $t_{\rm LJ} = 0.38$ ps, and thus frequency, and the units of force as $\epsilon_{0} / l_{0} = 21.9\,{\rm pN}$.

\section{Results} \label{results}
The objective of this study is to investigate the influence of various frequencies and different directions of the driving on the translocation times of the polymer. 

The chain is initially configured such that the first five monomers are inside the pore, while the remainder are aligned along the {\it cis} side along the $x$-axis in order to eliminate any possible complications regarding the geometry of the larger polymers, such as the formation of knots. During the thermalization phase of the simulations, which lasts for 1000 time units, the monomers within the pore are restricted in their movement. This thermalization time is long enough to reach an equilibrium state of the polymer in the {\it cis} region. Following this, the driving force is applied and the translocation process begins.

The Mean Translocation Time (MTT, also indicated as $\tau$ along the manuscript) is the primary measure of interest in this study. The TT is defined as the duration required for the final monomer of the polymer chain to enter completely inside the pore. In other words, the TT is the time at which the polymer has fully exited the cis side of the system. Figure~\ref{f-channel} shows the polymers while translocating inside the pore, under the application of the force ${\bf F}(t)$. Due to the stochastic nature of these simulations, the $\tau$ values are averages over a large number of experiments $N_{\rm exp}$, by changing the initial phase of the driving and random seed at every run. For the majority of the experiments $N_{\rm exp}=2000$, but for those with higher variance, and typically around some frequency of interest, we reached up to $N_{\rm exp}=10000$.

The longitudinal constant force in Eq.~\eqref{driving_f} $F_{\rm const}$ was introduced to promote the translocation even in the worst cases respect to the driving, {\it i.e.} opposite initial phase of the periodic force in which not all simulations result in a successful translocation. In these cases the polymer is not tethered to the pore, and in some realization the chains exit the pore remaining permanently on the {\it cis} side. These cases, whose number is highly reduced by the presence of $F_{\rm const}$, are not included in the MTT calculations.

\subsection{Transversal driving.} \label{results_trans}
In the transversal driving the force is applied in the $y$-axis, \emph{i.e.} $\hat{r} = \hat{y}$, with the functional form previously introduced in \eqref{Sinusoidal force}. The amplitude of the constant component of the force ($F_{\rm const}$) is kept at 0.3 so as not to obscure the effects of the transversal driving.

\subsubsection{Translocation time.}
\begin{figure}[]
 \centering
 \includegraphics[width=\linewidth]{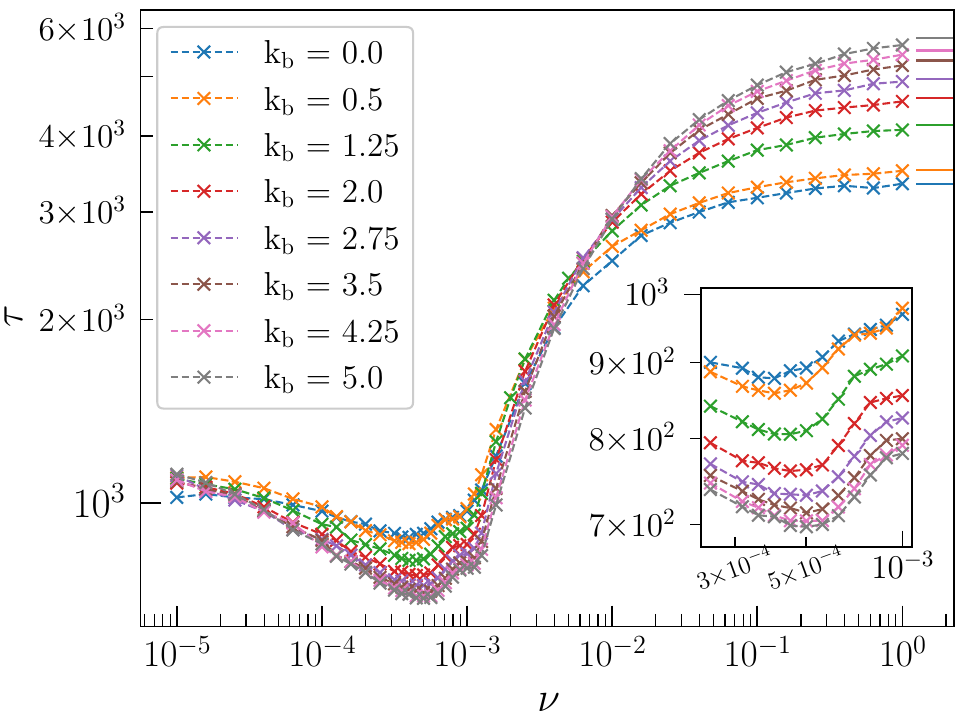}
 \caption{Mean translocation time as a function of the frequency of the transversal sinusoidal force applied for a chain of $ N = 30$, at different bending values, shown in the legend. The inset corresponds to the region where the minimum in translocation time is found. The continuous lines on the right correspond to simulations performed with the periodic driving off, representing the high frequency limit, and in clear agreement with the curves.}
\label{trans_sinusoidal}
\end{figure}

Figure \ref{trans_sinusoidal} displays the dependence of $\tau$ with the frequency of the driving force, as defined in equation \eqref{Sinusoidal force}, for a polymer chain consisting of $N = 30$ monomers and various bending constants, ranging from a fully flexible polymer ($k_{\rm{b}} = 0.0$) to an almost rigid rod ($k_{\rm{b}} = 5.0$). The graph reveals three distinct frequency regimes independently of the bending constant.

\vspace{0.2cm}
\emph{High frequency regime.} In this region, the TTs are considerably larger than those observed at lower frequencies, with an increase of almost an order of magnitude at the saturation. In this frequency range the period of the driving force is much smaller than the relaxation timescale of the polymer \cite{2016_Sakaue_escalado_pubmed, 2018_Menais_PRE_end_pulled}. Therefore, the transversal interaction cannot propagate effectively through the chain, and the force felt by the polymer results in a temporal mean of the sinusoidal excitation, {\it i.e.} a null contribution. This leads to a progressive descent of the effective intensity as the frequency increases, eventually reaching a point in which the transversal driving is practically 0, which corresponds to the observed plateau around $\nu = 10^{0}$. To confirm this, a set of simulations have been conducted with the driving force turned off, whose results have revealed a clear agreement with the higher values of the curves, and are plotted as continuous lines on the right of the curves.\\

\emph{Low frequency regime.} Here, the magnitude of the force throughout the entire translocation process is completely determined by its initial phase, since the period of the driving is much higher than the typical TT. 

The phase is uniformly distributed in the range $[0,2\pi)$, but the longness of the period makes the translocation occur during an almost fixed value of the oscillating force. It is worth to underline that, even if the cosine function averages over zero in a period, the contribution of the transversal driving to the polymers is in fact mediated by the interaction with the walls, which create a longitudinal force component of the same sign for the two cases: positive and negative values of the cosine function. This way, the transversal driving always aids the movement of the polymer towards the {\it trans} side, thus resulting in smaller TTs than those saturating at high frequencies.

\begin{figure}[]
 \centering
 \includegraphics[width=1.0\linewidth]{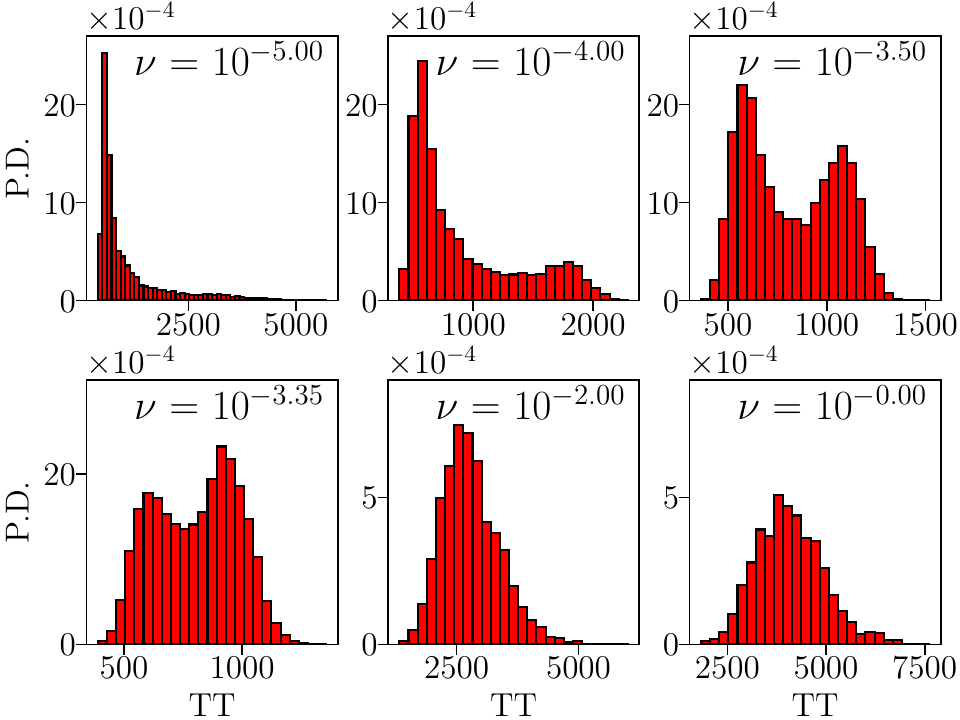}
 \caption{Histograms of the TTs at different frequencies for a chain of $N = 30$ and $k_{\rm{b}} = 1.25$. The frequencies, starting from the top left, are $\nu = 10^x , \; x = -5.0,\; -4.0,\; -3.5,\; -3.35,\; -2.0,\; 0.0$.}
\label{histogram_movement}
\end{figure}

\emph{Mid frequency regime.} This region is characterized by a clear decrease in the value of $\tau$s, expressed by the presence of a large frequency region with MTT values well below both the high and low frequency limits. This behavior is typical of the \emph{Resonant Activation} (RA) \cite{1992_Doering_PRE_Resonant_activation} effect that appears in the presence of an oscillating potential barrier accompanied by thermal fluctuations \cite{2010_Fiasconaro_Spagnolo_PB_resonant_activation, 2015_fiasconaro_falo_polymer_3D}. For this particular system, the RA effect has its origins in the interactions of the polymer with the walls in both the \emph{cis} and \emph{trans} sides of the membrane, giving rise to an effective potential barrier with oscillatory nature, caused by the periodic driving of the system. As a result the MTTs present a large minimum region of different order of magnitude with a modulation in the minimum proximity. The latter is due to the presence of classical oscillations caused by the deterministic periodic nature of the driving. These kind of modulations, that involve a smaller region of frequencies with respect to the RA effect, are also present in the absence of the potential barrier \cite{2010_Fiasconaro_Falo_PRE_mid_point}.

Fig. \ref{histogram_movement} provides the detailed features of the three frequency regimes previously discussed. The histograms show the distribution of translocation times recorded at different frequencies $\nu$ ranging from $10^{-5}$ to $10^0$. 

As already commented, in the low frequency region the translocation is dominated by the initial phase of the force. The low frequency peak therefore corresponds approximately to the maximum possible transversal excitation $F_{var} \approx \pm A$. This initial condition is more favourable than others due to choice of a uniform distribution of initial phases between 0 and $2\pi$, which give a non-flat distribution of the forces that depends on the cosine function, so making the extreme values of the forces more probable. Thus, the subsequent translocation times correspond to progressively smaller absolute values of the force, less represented in the probability distribution. 

Increasing the frequency leads to the mid frequency regime, which is characterized by the coexistence of two nearby peaks (See Fig.~\ref{histogram_movement} with $\nu\approx 10^{-3.35}$, $10^{-3.5}$ and $10^{-4.0}$) and a narrower TT distribution than those at low or high frequencies. In these conditions, the first peak corresponds to the minimum possible translocation time, \emph{i.e.} the translocation with the highest average force over the specific trajectory. The peak representing the fastest polymer translocation in the low frequency regime maintains its presence in the mid frequency region. Its disappearance marks the beginning of the high frequency regime.

The second peak corresponds to the translocation time of non-optimal initial conditions, \emph{i.e.} the TT for simulations in which the initial phase does not lead to a high value of the transversal force. Thus, the observed translocation time is higher than the time required to reach the optimum driving force giving the minimum TT, but small enough to allow the polymer to translocate at the subsequent high force values. This peak progressively moves towards lower values with increasing frequency, which can be understood in terms of the fact that the force period decreases.

At high frequencies ($10^{-2.0}, \; 10^{0.0})$, both of these peaks combine into a single, normal-like distribution. Additionally, the distribution moves towards higher TTs with increasing frequency as a consequence of the progressively lowering value of the effective periodic driving, eventually reaching 0 and leading to the plateau in the translocation times observed in Fig.~\ref{trans_sinusoidal}. 

\begin{figure}[]
 \centering
 \includegraphics[width=1.0\linewidth]{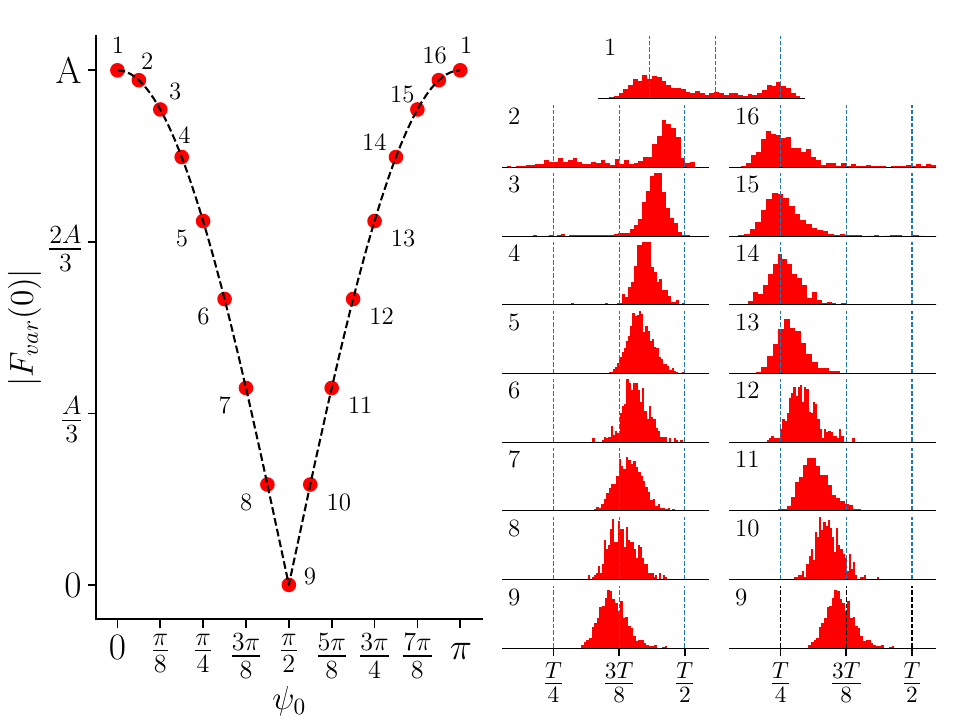}
 \caption{Absolute values of the force depending on the initial phase of the system in the range [0,$\pi$]. The histograms correspond to the TTs at the frequency minimum $\nu = 10^{-3.35}$, associated to the particular phase indicated by the index $i=1..16$, accompanied by three vertical dashed lines marking different fractions of the period: $T/4$, $3T/8$ and $T/2$. This graph corresponds to chains of length $N = 30$ and $k_{\rm b} = 1.25$}
\label{Phases_histograms}
\end{figure}

Looking back to the mid frequency regime, the relation between initial phase and the force value is further explored in Fig.~\ref{Phases_histograms}, which shows the distributions of the TTs for different $\psi_{0}$ values at the frequency of the minimum $\nu = 10^{-3.35}$. It is there evident that starting slightly before the maximum of the force ($\psi_{0} \sim 7\pi/8$, point 16 in the graph) results in the maximum average driving and therefore in the minimum possible translocation time. Conversely, the worst possible scenario occurs when starting at the maximum driving, where the chain either translocates in $T/4$ or need to wait, in average, an additional $T/4$ time lapse to reach the high forces again, resulting in a total translocation time of $\approx T/2$.

\begin{figure}[]
\centering
\includegraphics[width=0.95\linewidth]{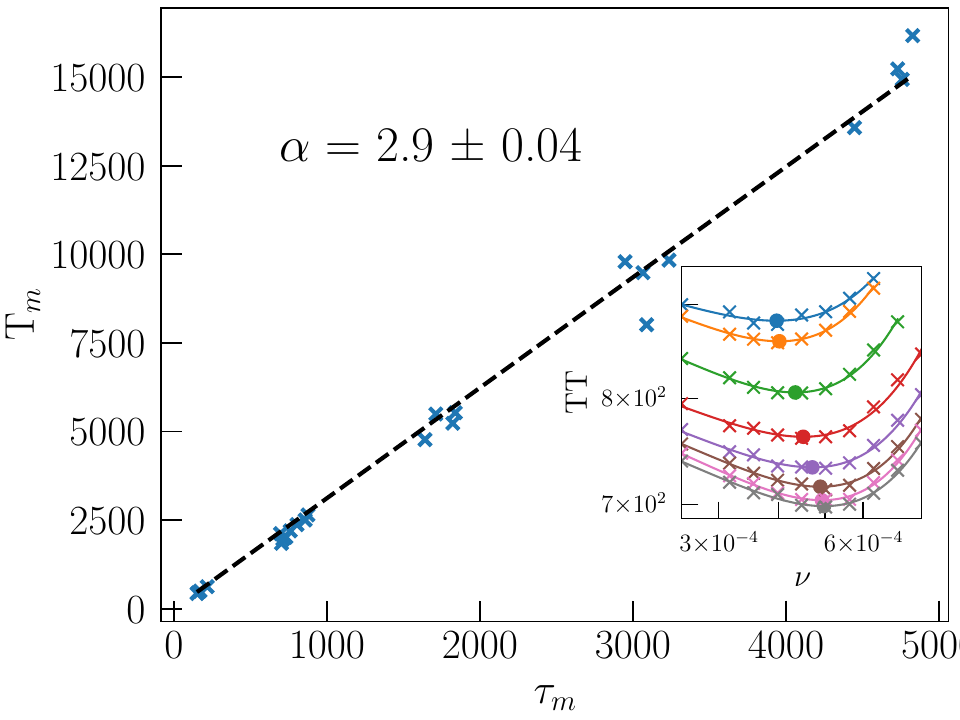}
\caption{Minimum translocation period as a function of the minimum translocation time for the different parameters used in the ms. The fitted line is shown in black, along with the parameter $\alpha$. In the inset: Parabolic fitting of the neighboring minimum points for the different bending curves of Fig. \ref{trans_sinusoidal}. Same color coding.} 
\label{parabolic_fitting}
\end{figure}

The intermediate initial phases result in intermediate translocation times, ranging from $T/4$ to $T/2$, making  the peak at $T/4$ the minimum possible $\tau$. The average over all the possible initial phases yields the translocation time distribution presented in Fig.~\ref{histogram_movement}. Chains with distinct flexibility constant $k_{\rm b}$s will exhibit distinct MTTs, but maintain the same condition for the minima: the minimum MTT, $\tau_{m}$, is found when the minimum possible translocation time (the TT with the maximum average driving) corresponds to $T/4$. As the distributions of translocation times at the minimum depend only on the initial phase of the system, a correlation between $\tau_{m}$ and its associated frequency period $T_{m}$ can be established for all chains, which results in the simple law:
\bea \label{linear_freq_tau}
    T_{m} = \alpha \tau_{m}
\eea
In fact, Fig.~\ref{parabolic_fitting}  collects the points of the minima taken for each set of parameters (chain length $N$, bending parameter $k_{\rm b}$, different amplitudes of the transversal driving $A$), where a fit with a parabolic equation around the minima has been performed in order to reduce fluctuations errors and the points of interest have been taken as the minima of the fit functions for all the parameters used: namely for the polymer lengths $N=30,\; 45,\; 60,\; 75$ and $k_b= 0.0, \;0.5, \;1.25, \;2.0, \;2.75, \;3.5, \;4.25, \;5.0$. The linear behavior of the points is evident and valid for all the parameters, with a slope of $\alpha = 2.90 \pm 0.04$. The points obtained with the fitting procedure are shown in the inset of Fig.~\ref{parabolic_fitting} with the corresponding fit curves for the chains of $ N = 30$ previously shown in Fig.~\ref{trans_sinusoidal}.

\subsubsection{Dependence on the bending parameter.}
Fig. \ref{trans_sinusoidal} contains several translocation curves corresponding to different values of the bending constant $k_{\rm{b}}$, showcasing the effect this parameter has on the TTs. 

At high frequencies, an increase of $\tau$ with $k_{\rm{b}}$ can be observed. This is attributed to the chain's conformation in the {\it cis} region, which is determined by that parameter. Chains with higher $k_{b}$ tend to be more extended in the {\it cis} region, while folded chains with lower gyration radius and closer to the pore are expected for low bending values. Therefore, the stiffer chains need to travel greater distances to achieve successful translocation, resulting in larger TTs \cite{2015_fiasconaro_falo_polymer_3D}.

This fact additionally explains the observed separation of the high frequency values. The persistence length, $L_{\rm p} \sim k_{\rm{b}}/k_{\rm B}T$, gives the length of the straight stick in which a chain can be divided, and corresponds in our case, due to the monomer distance equal to 1, to the number of monomers in that segment. Values of $k_{\rm{b}} = 0.0$ and $0.5$ have relatively small persistence lengths compared to the overall chain length ($L_{\rm p}=0$ and $L_{\rm p}=5$ respectively) and result in a similar behavior. On the other hand, for $k_{\rm{b}} = 1.25 \text{ or } 2.0$, ($L_{\rm p}=12.5$ and $20$) the persistence lengths are lower than --but comparable with-- the total chain length. Beyond $k_{\rm{b}} = 2.75$, $L_{\rm p}$ is either similar to, or larger than, the chain length, resulting in a saturating behavior of the MTT.

\begin{figure}[]
\centering
\includegraphics[width=0.95\linewidth]{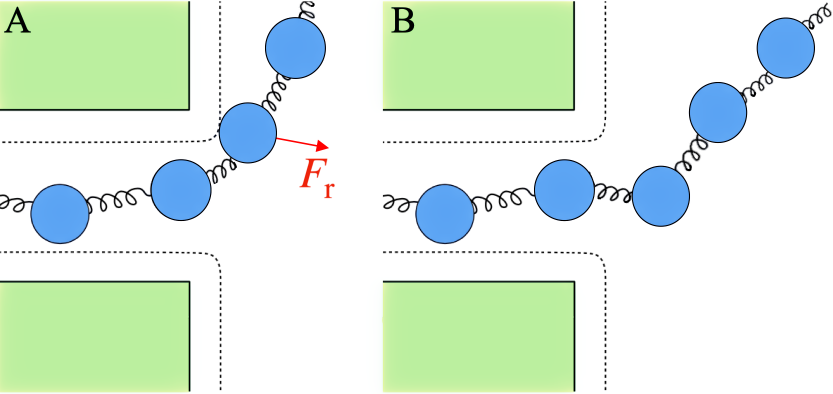}
\caption{Schematic depiction of the typical configurations found for chains with high (subfigure A) and low (subfigure B) values of $k_{\rm b}$. The term $F_{\rm r}$ corresponds to the reaction force that takes place on collision with the trans walls.} 
\label{wall_collisions}
\end{figure}

Around the minima this tendency fundamentally changes; chains with higher bending experience lower values of TTs, as the curves present a crossing interval around $\nu \approx 10^{-2}$. A possible explanation comes from the interactions of the chains with the walls on the trans side of the pore. The typical dynamics for chains translocating in the mid frequency range corresponds to the force starting at an already high value, and maintaining the same direction throughout the translocation process, leading to a lot of interactions with the trans walls. This reactions, which possess a longitudinal component, are what truly leads the chains forward, and they will be both stronger and more common the higher the $k_{\rm b}$ of the chain, given that those chains are more restricted in their possible conformations. This fact is illustrated in Fig.~\ref{wall_collisions}, that compares the typical geometries that lead to reactions of the walls (subfigure A) and an example of one that does not generate such forces (subfigure B). Chains with little to no bending will not collide with the {\it trans} walls as frequently and intensively as the more rigid ones, leading to more reaction forces and thus smaller translocation times for the higher $k_{\rm b}$ chains. This does not happen in the high frequency region because the effective transversal force eventually becomes 0, and the collisions with the trans walls are therefore reduced respect to the mid frequency regime. In general, the stiffness of the chain in the typical situations encountered in the mid frequency range gives an additional push in the longitudinal direction, thus explaining the lower MTTs for polymers with higher $k_{\rm b}$ \cite{2018_Sarabadani_end_pulled, 2010_Fiasconaro_Falo_PRE_mid_point}.

\subsubsection{Dependence on the polymer size N.}
\begin{figure}[]
 \centering
 \includegraphics[width=0.95\linewidth]{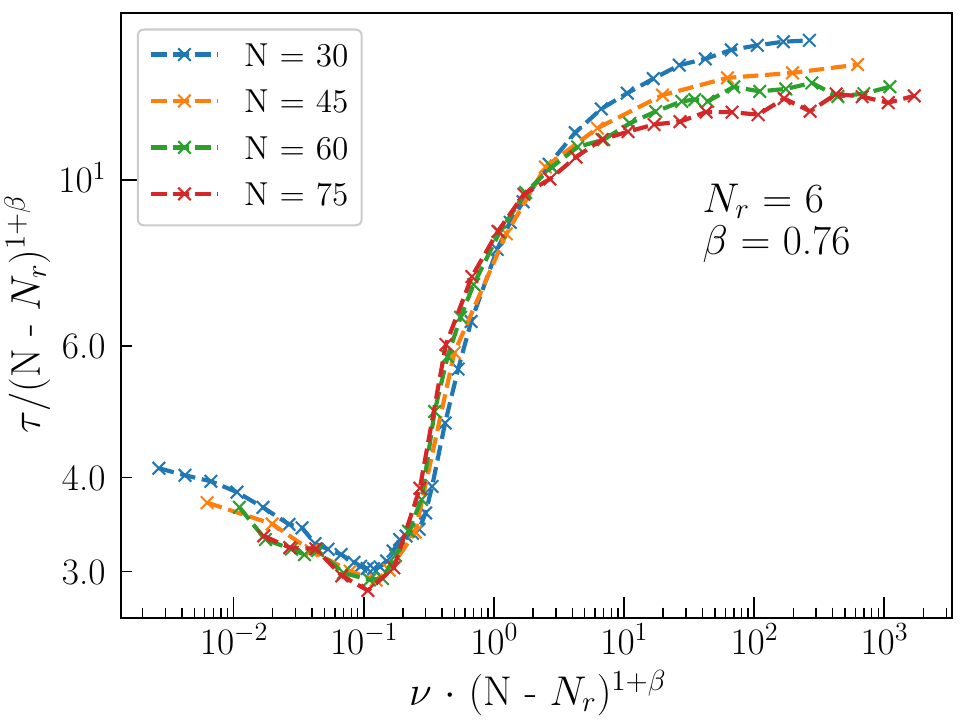}
 \caption{Transversal translocation time curves for chains with $k_{\rm{b}} = 1.25$ and different N, scaled according to equation \eqref{scale_law}.}
\label{scaling_transversal}
\end{figure}

As much as the polymer length increases, the translocation time also increases, and the minimum of the curves move towards lower frequency values, still maintaining the linear relation described by Eq.~\eqref{linear_freq_tau}. A scaling law for the MTTs of chains with different lengths $N$ can be established. The proposed form for this scaling is given by
\begin{equation}
    \tau \propto (N - N_{r})^{1 + \beta}
    \label{scale_law}
\end{equation}
where the parameter $N_r$ represents an effective reduction in the length of the polymers, caused by the confinement of a part of the chains inside the extended pore throughout the translocation process, and is around 5 to 6 monomers given the dimensions of the pore. The scaling coefficient $\beta$ is an exponent which depends on the bending constant of the chain $k_{\rm{b}}$. It is here shown that curves with the same $k_{\rm{b}}$ scales with $N$ according to Eq.~\ref{scale_law}, and the scaling parameters are shown in Table~\ref{scaling_table_1} Fig.~\ref{scaling_transversal} contains the scaled plot for chains of $k_{\rm{b}} = 1.25$ and different lengths.

Considering $N_r$ as a free parameter in the fit analysis, it is found that the best value for $N_{r}$ is $N_{r}\approx 6$ monomers, which confirms the effect of the finite pore extension on the translocation, defining an effective chain length.

\begin{table}[h]
\centering
\begin{tabular}{S |SSSS}
\hline
\hline
\textbf{$k_{\rm{b}}$} & 0.0 & 1.25 & 2.75 & 5.0       \\ \hline
\textbf{$\beta$}   & 0.59 & 0.76 & 0.82 & 0.92        \\
\textbf{$\sigma_{\beta}$} & 0.07 & 0.10 & 0.10 & 0.12 \\ 
\end{tabular}
\caption{Scaling exponent $\beta$ of the translocation times for the different bending parameters $k_{\rm b}= 0.0, 1.25, 2.75 {\rm and } 5.0$ in the transversal driving, and the corresponding standard deviation $\sigma _{\beta}$. In all cases $N_{\rm r}=6$. }
\label{scaling_table_1}
\end{table}

Previous determinations of the scaling coefficient $\beta$ have confirmed its relation with the Flory exponent $\beta_{F}$ \cite{2015_fiasconaro_falo_polymer_3D}, a parameter connected to the scaling  behavior of the radius of gyration ($R_g \sim N^{\beta_F}$) of a polymer chain, and that is also a function of the chain rigidity. For the flexible freely jointed chain model $\beta_{F}=0.5$, while the inclusion of the excluded volume interactions leads to $\beta_{F}\approx 0.6$ ($L_{p} \approx 0$) and a completely rigid rod ($L_{p} \gg N$) has $\beta_{F}=1.0$ \cite{1996_doi_introduction}. Our results show that all our coefficients belong to the appropriate range and increase monotonically with the bending constant, behaving consistently with the known properties of the Flory exponent.

\subsubsection{Analytical expression for low frequency values.}
At low frequencies, the amplitude and direction of the driving are determined by the initial phase of the system. This allows us to give an estimation of the mean translocation time of the system averaging over the different possible $\psi_{0}$. In order to make these calculations easier to manage, let us consider a square wave, which roughly corresponds to the sinusoidal force excitation approximated to 0 or its extreme values $\pm A$. It is worth to remember that both extreme values of the force aid the translocation in the same manner, given that the driving is applied in the transversal direction. The evolution of this square wave and the comparison with the sinusoidal interaction (in terms of module) can be seen in the appendix, in Fig.~\ref{analytical_trans_depict}. 

With that description of the square wave, there are only two timescales of relevance: $\tau_{\rm on}$, the time taken by the chain to translocate if the driving is in its extreme values $\pm A$, and $\tau_{\rm off}$, \emph{i.e.} the time it takes when the transversal force value is 0 during the process. We then have four distinct translocation possibilities, for a given initial phase: either the force starts and remains at the same value, or we will experience a switch between high and low force, or viceversa. Given that, the formula can be written as (see the Appendix for an extended derivation of this formula):
\bea
    \langle TT \rangle &=& \frac{2}{3} \tau_{\rm on} \left(1 - \frac{\tau_{\rm on}}{T/3}\right) + \frac{1}{3} \tau_{\rm off} \left(1 - \frac{\tau_{\rm off}}{T/6}\right) +      \\ 
    &+& \frac{2}{3} \frac{\tau_{\rm on}}{T/3} \left(\frac{\tau_{\rm on}}{2} + \frac{\tau_{\rm off}}{2}\right) + \frac{1}{3} \frac{\tau_{\rm off}}{T/6} \left(\frac{\tau_{\rm on}}{2} + \frac{\tau_{\rm off}}{2}\right) \nonumber 
    \label{low_freq_analytical}
\eea
Note that the above derivation requires the assumption that both $\tau_{\rm on}$ and $\tau_{\rm off}$ are smaller than the time spent by the driving in the corresponding values ($\pm A$ or $0$), which is $T/3$ and $T/6$, respectively. Since $\tau_{\rm off}$ has to be larger than $\tau_{\rm on}$, this gives an upper limit for the frequency range in which the approximation remains valid:
\be
    \tau_{\text{off}} \leq \frac{T}{6} \; \rightarrow \; \nu \leq \frac{1}{6\tau_{\text{off}}}.
\label{tau_limit_trans}
\ee
\begin{figure}[H]
 \centering
 \includegraphics[width=0.95\linewidth]{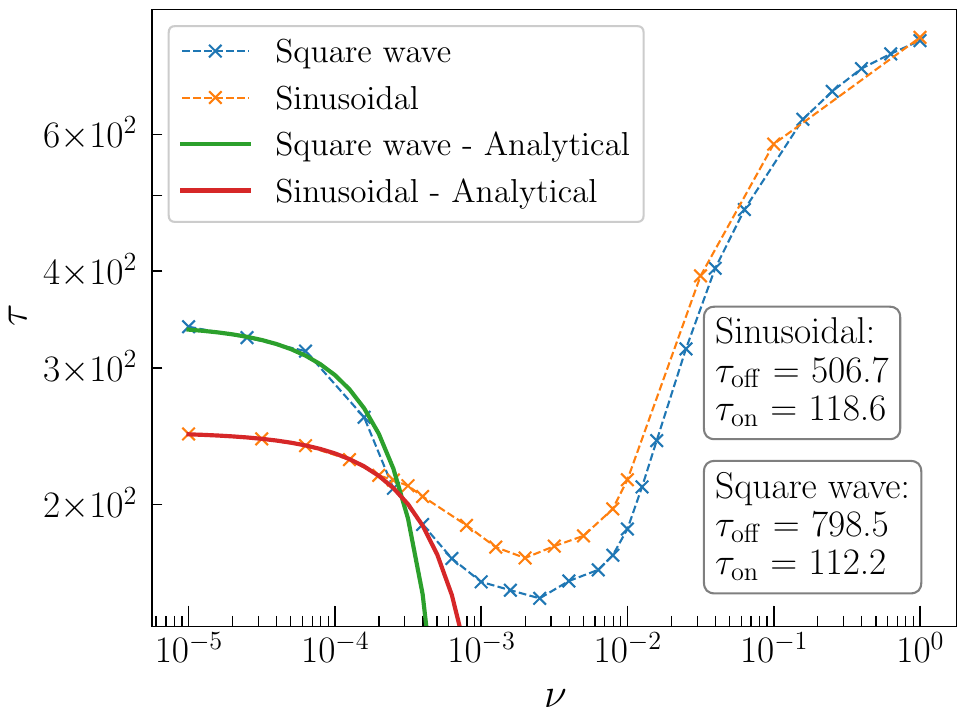}
 \caption{Translocation curves for polymer chains of $N = 15$ and $k_{\rm{b}} = 1.25$, with square wave and cosinusoidal drivings, along with their low frequency analytical approximations \eqref{low_freq_analytical}. The values of the parameters are included in the graph.}
\label{square_comparison}
\end{figure}

The comparison of the analytical derivation with the simulations are shown in Fig.~\ref{square_comparison} for a chain of $N = 15$ and $k_{b} = 1.25$. The blue points are the simulations of the MTTs with the square wave, and the continuous line is the analytic approximation, which results in an excellent agreement at low frequency. The parameters used, $\tau_{\rm on} = 112.16$ and $\tau_{\rm off} = 798.49$, have been obtained from simulations performed at maximum ($\pm A$) and minimum ($0$) driving, respectively.

The analogue approximation of the sinusoidal driving, has been also included in Fig.~\ref{square_comparison}, with, again, an excellent agreement between formulas ennd simulations. In this case, the parameters $\tau_{\rm on}$ and $\tau_{\rm off}$ come from a fitting procedure. Both analytical expressions fail on the hard limit established in Eq.~\ref{tau_limit_trans} at the value $\nu_{\rm lim} \approx 10^{-3.68}$.

\subsection{Longitudinal driving}\label{results_long}
The longitudinal driving define the application of the periodic force along the pore, {\it i.e.} along the $x$-axis ($\hat{r}=\hat{x}$), with the same temporal dependence indicated in Eq.~\eqref{Sinusoidal force}. 

Things here are different than in the transversal case, where the introduction of the periodic force leads to a general descent of the translocation times whatever frequency is considered, given that the application of the force, though oscillating sinusoidally, always aid the translocation process, as it contributes with the $x$-component of the wall reactions. That symmetry is broken in the case of the longitudinal driving, where the negative values of the cosine, give forces opposing the translocation movement. In order to ensure that we have translocating trajectories in the majority of the simulations performed, the constant term of the force ($F_{\rm const}$) has been then increased in value, going from the original 0.3 up to 1.8 in program units.

\subsubsection{Translocation time.}
The evolution of the average translocation times with the frequency on the longitudinal driving can be found in Fig.~\ref{long_sinusoidal}, containing chains of $N = 30$ with different values of the bending constant $k_{\rm b}$. Just as with the transversal driving, three distinct frequency regimes can be identified:

\begin{figure}[]
 \centering
 \includegraphics[width=\linewidth]{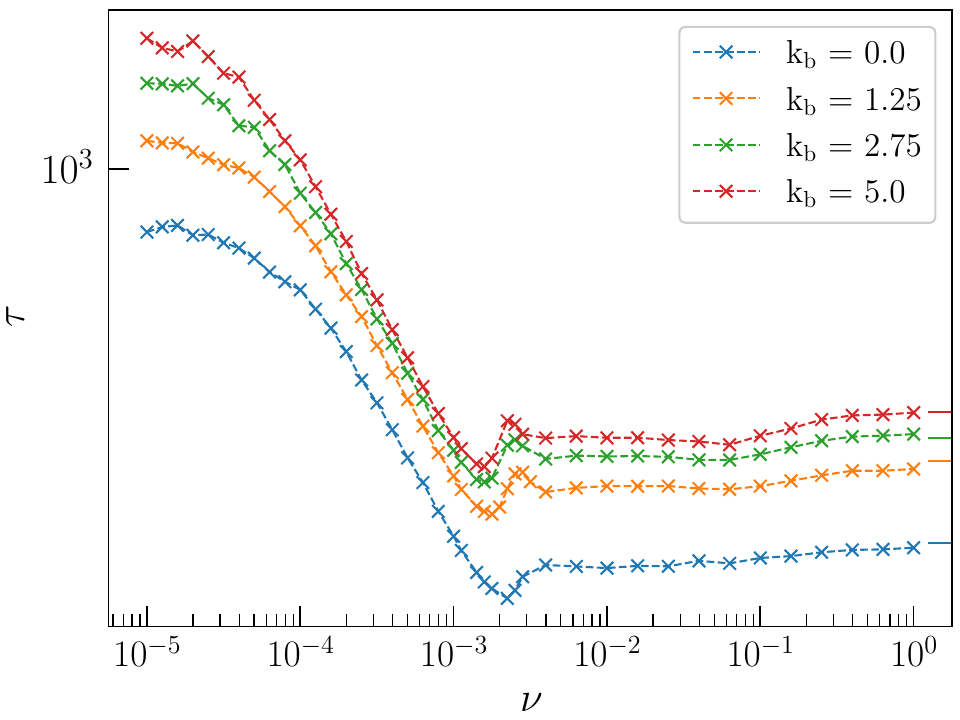}
 \caption{Translocation time as a function of the frequency of the longitudinal sinusoidal force applied for a chain of $ N = 30$, at different bending values, shown in the legend. The continuous lines on the right correspond to the MTTs  when the periodic driving is off, pointing out the saturation of the MTTs in the high frequency limit.}
\label{long_sinusoidal}
\end{figure}

\emph{Low frequency regime.} In this region, as seen in the transversal driving, the magnitude of the force is defined by the value of the initial phase. In contrast to the transversal case, the translocation times at low frequency are much higher than those at high frequency. This is due to the absence of the symmetry previously present in the transversal driving when passing from positive to negative values of the cosine: negative values of the force, which aided the translocation in the transversal scenario, now oppose to it, increasing the translocation times with respect to the positive forces. Thus, while in the transversal case the TTs are smaller that the high frequency values, in this longitudinal case the average translocation time is much higher than in the high frequency because it is greatly increased by the effect of negative forces.

\emph{High frequency regime.} At high frequencies the translocation times eventually saturate. This happens for period of the forces much smaller than the polymer relaxation timescale \cite{2016_Sakaue_escalado_pubmed, 2018_Menais_PRE_end_pulled} and, similarly to what we found in the transversal case, the oscillations of the force average to zero so that  the MTT only depends on the constant contribution $F_{\rm const}$. Simulations with the time-dependent force turned off, leaving only the constant term, can be seen in the right of the graph as continuous lines, in clear agreement with the data and showing the saturation of the MTTs.

\emph{Mid frequency regime.} In the mid frequency range a clear decrease of the translocation times is appreciated, in combination with a series of oscillations around the minimum point of the curves. 
The results are in all similar to those already obtained with pore driven forces (See \cite{2011_Fiasconaro_Spagnolo_RA, 2015_fiasconaro_falo_polymer_3D, 2013_Ikonen}), and the results of the transversal driving. Similarly, the non monotonic behavior corresponds to the combination of two phenomena: the Resonant Activation (RA) effect given by the interaction of the polymer with the walls in both sides of the pore\cite{2010_Fiasconaro_Spagnolo_PB_resonant_activation, 2015_fiasconaro_falo_polymer_3D, 2013_Ikonen}, and a classical resonant effect given by the synchronization of the polymer chain with the driving frequency, responsible of the oscillations \cite{2010_Fiasconaro_Falo_PRE_mid_point}.

\begin{figure}[]
 \centering
 \includegraphics[width=\linewidth]{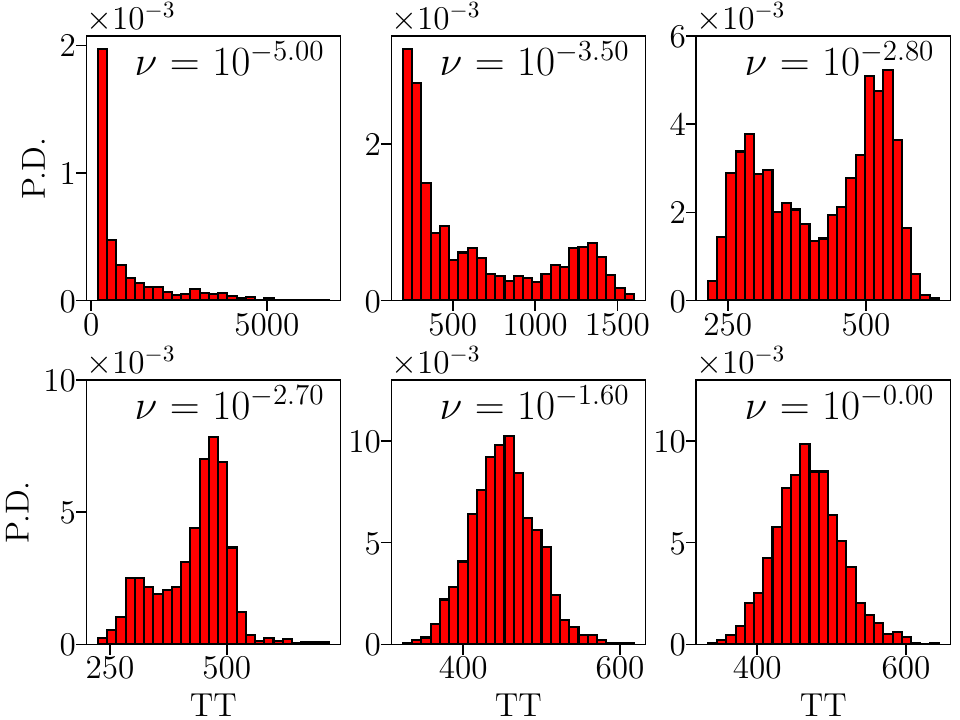}
 \caption{Histograms of the TTs at different frequencies for chains of $N = 30$ and $k_{\rm b} = 1.25$. The frequencies, starting from the top left, are $\nu = 10^x , \; x = -5.0,\; -3.5,\; -2.8,\; -2.7,\; -1.6,\; 0.0$.}
\label{histogram_long}
\end{figure}

The histograms of the mean translocation times at several values of the frequency, from $10^{-4}$ up to $10^{0}$, are presented in Fig.~\ref{histogram_long}.

At low frequencies, where the value of the force along the trajectory is approximately fixed during the dynamics and strongly determined by the initial phase, we observe a high peak at low values of the TTs, corresponding to translocations characterized by force values around $+A$ \emph{i.e.} the minimum possible translocation time, followed by a long tail, due to other initial phases. This tail is due to the negative value of the periodic force, which gives much larger translocation times and a distribution with higher variance.

High frequency histograms are described by normal-like distributions centered around the translocation time given by the constant term of the force, due to the value of the periodic driving being effectively 0. The distribution progressively shifts towards higher force values as the frequency increases, eventually arriving at a saturation value, corresponding to the constant force indicated by the straight lines in Fig.~\ref{long_sinusoidal}.

In the mid frequency regime two nearby peaks can be observed. The first corresponds to the minimum possible translocation time, and takes place when the averaged periodic force the chain experiments is maximal. This peak remains at its position from the start of the low frequency regime, eventually disappearing after combining with the second peak, marking the beginning of the high frequency region.

The second peak corresponds to the average of the translocation times of the different chains that do not start at high enough values of the force. Therefore, it is a combination of the time required to reach high forces and the time necessary to translocate with those forces. It moves towards lower time values with increasing frequency, because the time necessary to reach high forces diminishes, as a consequence of the decreasing period of the driving.

\begin{figure}[]
 \centering
 \includegraphics[width=\linewidth]{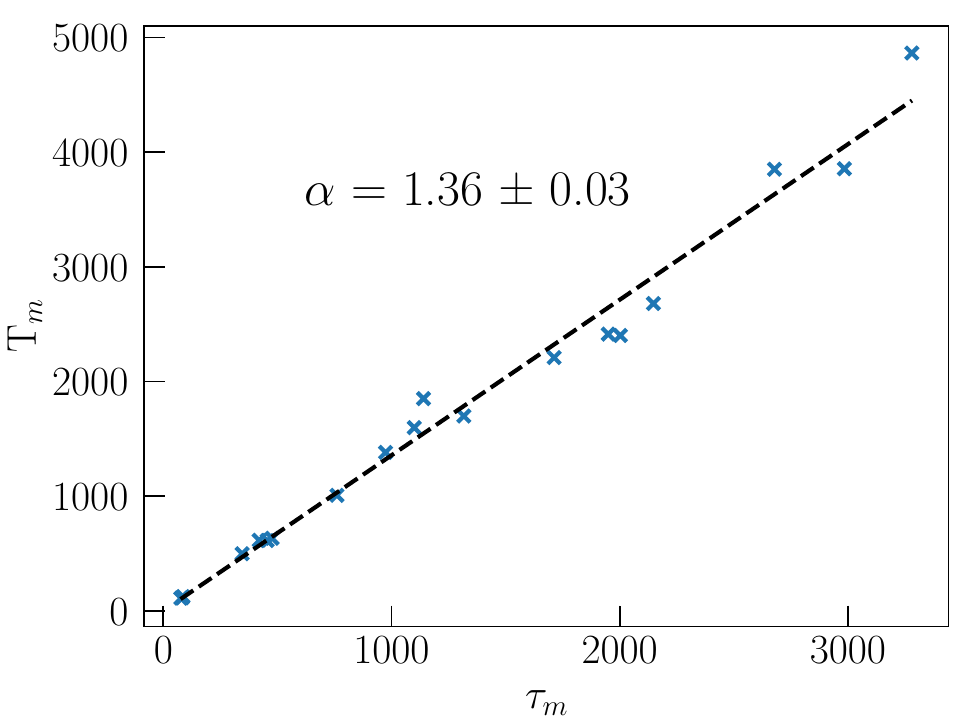}
 \caption{Minimum translocation period as a function of the minimum mean translocation time for the different parameters considered. The fitted line is shown in black, along with the value of $\alpha$.}
\label{alpha_longitudinal}
\end{figure}

The minimum in the translocation curves is reached when the value of the first peak (corresponding to the minimum possible TT, associated to maximum driving during the translocation process) corresponds to a half of the driving period $\tau_{\rm min} = T/2$. Once again, independently of the features of the chains such as length or rigidity, a linear relationship between the minimum average translocation time and the period associated to it can be established, like in \eqref{linear_freq_tau}. Different minimum MTT $(\tau_{m})$ and minimum period ($T_{m}$) pairs for chains with different values of number of monomers $N$ and bending constant $k_{\rm b}$ are contained in Fig.~\ref{alpha_longitudinal}, along with the fit for the $\alpha$ coefficient and obtaining $\alpha = 1.36 \pm 0.03$, in agreement with previous studies \cite{2010_Fiasconaro_Falo_PRE_mid_point}.

\subsubsection{Dependence on the bending parameter.}
For the longitudinal translocation the MTTs monotonically increase with the bending constant $k_{\rm b}$ of the chain, as can be seen in Fig.~\ref{long_sinusoidal}, where the value of the bending $k_{\rm b} = 0.0,\; 1.25,\;2.75 \text{ and } 5.0$ are shown. This difference can be explained from the typical conformations the chains tend to adopt depending on their $k_{\rm b}$ values along the translocation process. In the {\it cis} region, more rigid chains will tend to be in extended conformations along the axis of the pore. This means that their mass centers will be further from the pore entrance respect to more flexible change and will therefore have to travel longer distances to achieve successful translocation. This hypothesis is further confirmed by the fact that the differences between the chain with $k_{\rm b} = 0$ and the rest are significantly larger than those between the chains of $k_{\rm b} = 1.25$ and above, which have persistence lengths ($L_{\rm p} \approx 10k_{\rm b}$) either comparable or larger than the chain length $N = 30$ in Fig.~\ref{long_sinusoidal} and thus, their conformations will be similar.

When applying the transversal driving, we observed an inversion of this tendency around the mid frequency range, presumably associated to the differences in interactions with the walls between chains with different $k_{\rm b}$ values. However, for the longitudinal driving, that is not the case. Interactions with the \emph{trans} walls with the longitudinal driving are evidently weaker and less frequent than in the transversal driving, given that all the forces are concentrated in the $x$-axis, and the tendency established at high frequencies is maintained throughout all the frequency range. 

\subsubsection{Dependence on the polymer size N.}
Increasing the length of the chains leads to higher values of the translocation times, and displacements of the translocation minima towards lower frequency values in order to verify the linear relation between the minimum translocation time and its associated minimum period.

We used the law previously established in Eq.~\eqref{scale_law} to scale the values of the translocation times and frequencies for chains of different lengths $N$ and the same bending constant. 
\begin{equation}
    \tau \propto (N - N_{r})^{1 + \beta}
\end{equation}
where $N_{r}$ represents, again, a reduction in the effective number of monomers of the chain resulting from the extended nature of the pore, so that the movement of the chain is caused by the remaining part of the chain, and $\beta$ is the scaling exponent that depends only on the bending constant $k_{\rm b} $ of the chain. Translocation times and frequencies of the polymers of different lengths $N$ and equal bending constants were rescaled according to this law, fitting the values of the parameter to those at which the curves collapsed against each other. A value of $N_{\rm r} \approx 6$ was obtained, just as in the transversal driving, so confirming its pore estended relationship. The values of the scaling parameter $\beta$ are listed in Table.~\ref{scaling_table_2}, together with the standard deviation related to its determination.

\begin{table}[h]
\centering
\begin{tabular}{S |SSSS}
\hline
\hline
\textbf{$k_{\rm{b}}$} & 0.0 & 1.25 & 2.75 & 5.0       \\ \hline
\textbf{$\beta$}   & 0.64 & 0.73 & 0.79 & 0.83        \\
\textbf{$\sigma_{\beta}$} & 0.02 & 0.02 & 0.03 & 0.03 \\ 
\end{tabular}
\caption{Scaling exponent $\beta$ of the translocation times at the longitudinal driving for the different bending parameters $k_{\rm b}=0.0, 1.25, 2.75, 5.0$, and the corresponding standard deviation $\sigma_{\beta}$. In all cases $N_{\rm r}=6$.}
\label{scaling_table_2}
\end{table}

\begin{figure}[]
 \centering
 \includegraphics[width=0.95\linewidth]{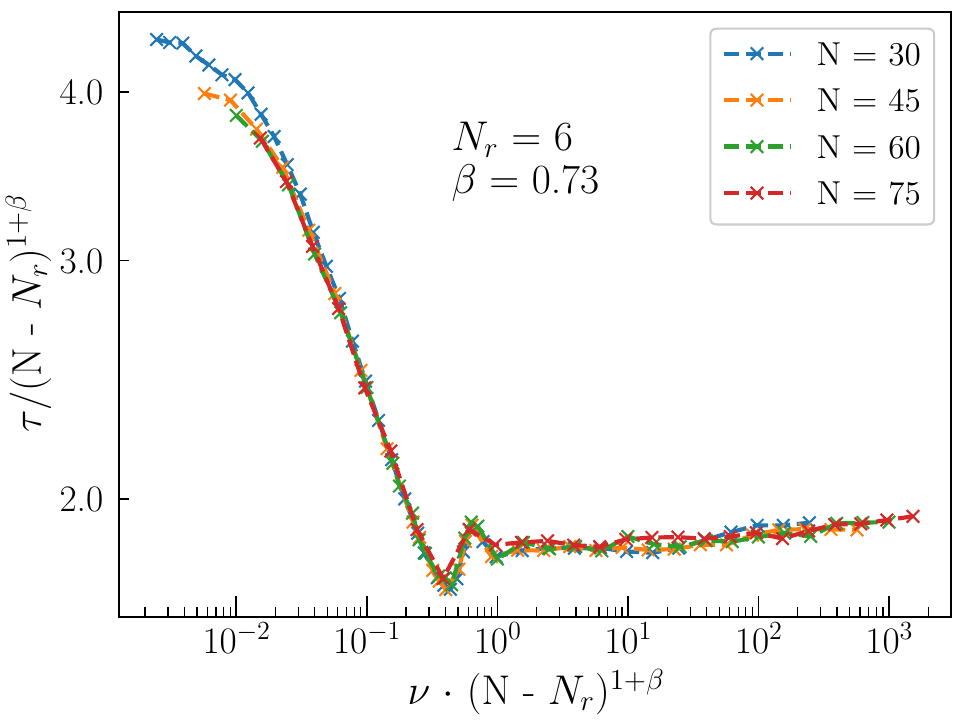}
 \caption{Longitudinal translocation time curves for chains with $k_{\rm{b}} = 1.25$ and different N, scaled according to equation \eqref{scale_law}.}
\label{scaling_longitudinal}
\end{figure}

The coefficients obtained are within the range of the Flory exponent, and monotonically increase with the value of the bending constant $k_{b}$, as expected. The overlapping of the  scaled curves corresponding to chains of $k_{\rm b} = 1.25$ and $N = 30,\; 45,\; 60,\; 75$ can be seen in Fig.~\ref{scaling_longitudinal}.

\subsubsection{Analytical expression for low frequency values.}
Following the same ideas previously explained in the transversal driving, and given that at low frequencies the driving is completely determined by the initial phase, an analytical expression for the translocation curve can be derived when averaging over the different initial phases $\psi_{0}$. Once again let us consider a square wave, with extreme values $\pm A$, which, in principle can approximated a sinusoidal dependence.

For the traversal driving, we dealt with two timescales, given the equivalence between the extreme $+A$ and $-A$ in terms of translocation times. However, that no longer holds for the longitudinal driving. Let us then consider three different timescales: $\tau_{\rm +}$, the time taken for the translocation to complete when the force value is $+A$, $\tau_{\rm 0}$, the time it takes when the driving is 0, and $\tau_{\rm -}$, the time necessary to translocate when the amplitude of the force is at its lowest point, $-A$. Given that we now have 3 timescales, and the functional form of the square wave as depicted in Fig.~\ref{analytical_long_depict}, we have now 7 scenarios to consider, depending on the initial phase: either staying in the same force during the trajectory or changing between two of those values. Taking into account the possibility of starting into each of the regimes, and the probabilities of changing between them, the formula yields (see the appendix for a detailed derivation)
\bea 
    \langle TT \rangle = \frac{1}{3} \tau_{\rm on} \left(1 - \frac{\tau_{\rm on}}{T/3}\right) + \frac{1}{3} \tau_{\text{off}} \left(1 - \frac{\tau_{\text{off}}}{T/3}\right)\nonumber\\ + \frac{1}{3} \tau_{0} \left(1 - \frac{\tau_{0}}{T/6}\right) + \frac{1}{3} \frac{\tau_{\rm on}}{T/3} \left(\frac{\tau_{\rm on}}{2} + \frac{\tau_{0}}{2}\right)\nonumber\\ 
    + \frac{1}{3} \frac{\tau_{\text{off}}}{T/3} \left(\frac{\tau_{\text{off}}}{2} + \frac{\tau_{0}}{2}\right) + \frac{1}{3} \frac{\tau_{0}}{T/6} \left(\frac{\tau_{0}}{2} + \frac{\tau_{\rm on}}{4} + \frac{\tau_{\text{off}}}{4}\right).
    \label{analytical_long_2}
\eea

\begin{figure}[]
 \centering
 \includegraphics[width=0.95\linewidth]{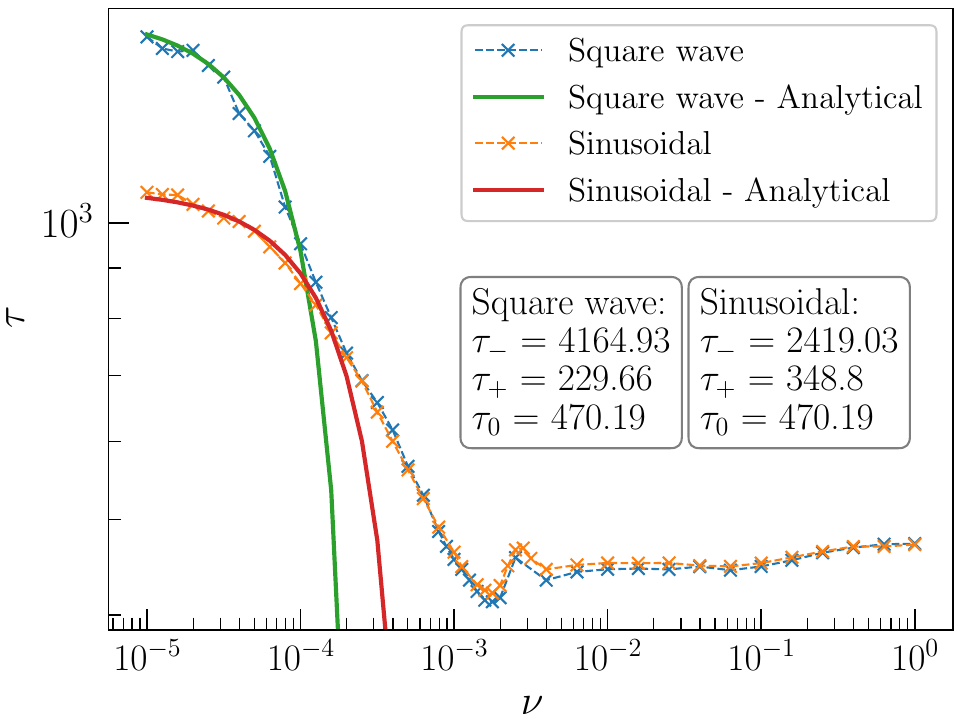}
 \caption{Translocation curves for polymer chains of $N = 30$ and $k_{\rm{b}} = 1.25$, with square wave and sinusoidal drivings, along with the fit of their low frequency analytical functions (Eq.~\eqref{analytical_long_2}). The legends show the values of the parameters used for the two cases and calculated as discussed in the text.}
\label{square_comparison_long}
\end{figure}

Also for the longitudinal case it is possible to establish a maximum frequency limit for the validity of the above formula. For the probabilities to be contained between 0 and 1, we must ensure that $\tau_{+} \leq T/3$, $\tau_{0} \leq T/6$ and $\tau_{-} \leq T/3$. Either $\tau_{0}$ or $\tau_{-}$ will give the limiting condition of the system. In our particular case, given the numerical values of the amplitude of the force $A = 1.5$ and the constant term $F_{const} = 1.8$, we will likely have $\tau_{0} \gg \tau_{-} $, and $\tau_{-}$ will be the limiting factor. The frequency limit is then 
\bea
    \tau_{-} \leq \frac{T}{3} \; \rightarrow \; \nu \leq \frac{1}{3\tau_{-}}.
    \label{frequency_limit_longitudinal}
\eea
In order to validate the analytical expression, a set of simulations have been run with the longitudinal driving considering both a square wave and a sinusoidal excitation. The results can be visualized in Fig.~\ref{square_comparison_long}, where a perfect agreement of the curves over the data are plotted. The blue points correspond to the square wave, with the green continuous line being its associated analytical curve, the parameters of which can be seen in the figure. These values for the three timescales have been calculated from simulations performed at the different possible amplitudes of the force $+A$, $0$ and $-A$, and then used to plot the analytical curve of the square wave driving.

The orange points and the red line correspond to the simulations and analytical approximations of the sinusoidal excitation, respectively. Both the $\tau_{+}$ and $\tau_{-}$ parameters have been here fitted with Eq.~\eqref{analytical_long_1}, by leaving $\tau_{0}$ as calculated in the square wave simulations.

Considering the value of $\tau_{-}$ for the square wave, the formula begins to fail at $\nu \approx 10^{-4.09}$.

\subsection{Comparison between longitudinal and transversal drivings.}

\subsubsection{Effect on the translocation times.}
\begin{figure}[]
 \centering
 \includegraphics[width=0.95\linewidth]{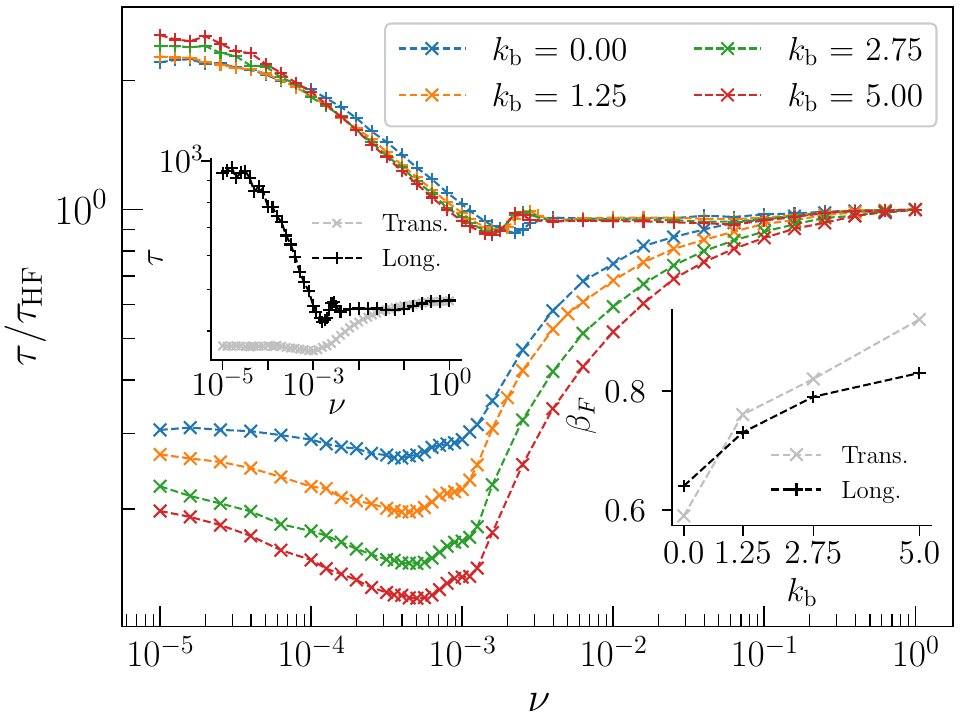}
 \caption{Main plot: Translocation curves for polymer chains with $N = 30$ and different $k_{\rm{b}}$, for both the transversal (bottom) and longitudinal (top) regimes. The TTs are normalized with their corresponding high frequency values ($\tau_{\rm HF}$) to eliminate the effect of the different constant terms of the force used. Left inset: example of the comparison between the non normalized TTS for the two drivings with $N=30$ and $k_{\rm b}=1.25$ with the common saturation trend at high frequencies. Right inset: comparison of the $\beta_{\rm F}$ values presented in Table~\ref{scaling_table_1} and Table~\ref{scaling_table_1}: The scaling coefficient with the chain length tends to 1 (straight stick) easier under the transversal driving than under the longitudinal.}
\label{comparison_trans_long}
\end{figure}

Both of the drivings considered in this work exhibit a global minima in the translocation times at a given frequency of the oscillating force. However, their behaviors are completely different. Fig.~\ref{comparison_trans_long} contains the curves for both longitudinal and transversal driving, normalized by their high frequency values.

As it can be seen, their behaviors appear very different; both drivings generate a saturation trend for both regimes: low and high frequencies. And both drivings produce the large minimum region that reveals the RA optimization. However, the relative behavior for each of the two driving appears very different, as the extreme values are opposite in value: higher at high frequency than at low ones for the transversal driving, and lower at high frequency for the longitudinal driving. This different behavior is due to the fact that the presence of the transversal driving always acts in favour of the translocation, due to the {\it trans} walls reaction which always act in the $x$-direction, so that the highest $\tau$ value is reached for high frequency where the transversal force mediates to zero. Conversely, in the longitudinal case the translocation is either helped or hindered depending on the phase of the cosine, so resulting in a general increase of the MTT at low frequencies with respect to the saturating behavior at high ones.

The different behavior with $k_{\rm b}$ can be appreciated from Figs.~\ref{trans_sinusoidal}~and~\ref{long_sinusoidal}. The longitudinal driving exhibits a monotonic dependence with the bending parameter; chains with higher $k_{\rm b}$ present higher TTs on average, the plausible explanation being given by the typical conformation of the chain after thermalization: more rigid chains will tend to be elongated, implying that they will have to traverse a higher distance to achieve translocation.

For the transversal driving this tendency varies with frequency, presumably due to the interactions with the walls. They are more relevant in the transversal driving, given that the periodic force is applied in the $y$-axis. At high frequencies, since the periodic force felt by the chain averages to 0, the tendency from the longitudinal case is recovered, whereas for the mid frequency the opposite occurs: chains with higher $k_{\rm b}$ exhibit lower TTs, given that the interactions of the walls will be less common for the more rigid polymers. At low frequency, as mentioned before, a clear tendency cannot be derived for the transversal driving.

Regarding the scaling with the length of the chains $N$, both drivings show a similar tendency; the effect of the pore, given by $N_{r}$ in Eq.~\eqref{scale_law} is the same in both regimes, while the $\beta$ coefficients, although similar, differ in value, although following the monotonic tendency and limited to the range characteristic of the Flory exponent. The comparison between both, showing the $\beta_{F}$ values for the different $k_{\rm b}$, can be found in the right inset of Fig. \ref{comparison_trans_long}.

\subsubsection{Differences between the $\alpha$ values.}
The absolute minima of the TTs present a clear linear relation with its associated period for both the longitudinal and the transversal driving, the slope being encoded in the parameter $\alpha$. The origin of this linear relationship is rooted in the condition for finding the existence of the minima: it appears when a fraction of the period, $T/4$ for the transversal and $T/2$ for the longitudinal, is equal to the minimum possible translocation time of the chain, \emph{i.e.} in the condition in which the chain translocates with during the maximum driving. In that minima, the histogram of the translocation times for the chain presents two peaks: one at $T/4$ (or $T/2$), and another at a higher value of the translocation times, which results from the average over all the initial phases that do not lead to translocation with the minimum time, as shown in Fig.~\ref{Phases_histograms}. Making a rough approximation of the peaks, they situate themselves next to $3T/8$ in the transversal, and $3T/4$ in the longitudinal.  

The timescales for the longitudinal driving seem to be double of those involved in the transversal, since in the former the negative force when applied on the $x$-axis does not aid the translocation like its positive counterpart, as it happens in the latter case. With this in mind, we expect the coefficient $\alpha$ in the longitudinal driving be one half of that one in the transversal case. In fact, comparing the values, we find:
\[
    \alpha_{\rm long} = 1.36 \pm 0.03 \approx 1.44 \pm 0.02 = \frac{\alpha_{\rm trans}}{2},
\]
which seems to closely verify our assumption. Possible errors may be related to either the calculation of the minima themselves, or point out the need for more simulations of our system.

\section{Summary and Conclusions.} \label{conclusions}
In this work we study the translocation of a polymeric chain through a extended pore under the action of A periodic driving applied in two directions respect to the pore axis: transversal and longitudinal. We studied the effect that the different force frequencies exert on the MTTs, as shown in Fig.~\ref{trans_sinusoidal}~\&~\ref{long_sinusoidal}, finding three different regimes of behavior depending on the frequency values: the high frequency range, in which the periodic driving averages effectively at 0, the low frequency range, in which the translocation forces are determined by the initial phase of each simulation and the mid frequency range, in which a global minimum in the translocation times is present.

The curves point out specific differences between the two drivings. The choice of the transversal case immediately leads to smaller MTTs than the high frequency limit, whereas for the longitudinal driving the MTT can be higher or lower depending on the frequency, observing a decrease in the mid frequency region but an increase at low frequencies, due to the phase giving negative values of the time-dependent force.

Regarding the minimum region, it is obtained as the combination of the Resonant Activation effect, originating from the interactions of the polymer with the walls of the pore, and classical oscillations given the periodic nature of the driving. We found that the minimum for the transversal and the longitudinal cases takes place when the frequency of the force verifies $\tau_{min} = T/4$ or $\tau_{min} = T/2$, respectively, with $\tau_{min}$ being the minimum mean translocation time possible, \emph{i.e.} the average TT at maximum driving values. This clear dependence allows for the possibility of establishing a linear relationship between the TTs in the minimum and its associated period, with a proportionality factor $\alpha$, different for our two force directions, but independent of the physical parameters of the chains.

Different behavior between the driving types when changing the bending constants $k_{\rm b}$ has been observed. The longitudinal driving exhibits a monotonous dependence on the bending parameter, with higher bending leading to higher MTTs, whereas for the transversal driving the effect of $k_{\rm b}$ was different in each of the frequency regimes. This different behavior may have its origin in the interactions with the walls of the pore in the {\it cis} and {\it trans} side of the membrane, more probable in the transversal translocation given that the force is not applied in the $x$-axis.

A scaling law for different lengths of the chain is also verified, with the form $(N-N_{r})^{1+\beta}$, where $\beta$ corresponds to the Flory exponent for semiflexible chains. The fit parameters for this scaling reveals the effect of the pore length on the translocation, revealing the effective number of monomers inside the pore $N_{r}$, and imposing an effective length of the chain, while the $\beta$ coefficient is comprised in the limits of the Flory exponent and increases monotonically with the bending constant $k_{\rm b}$.
Finally, an analytical expression for the low frequency values is calculated for both drivings with an excellent agreement between the simulation and the analytical functions. It is correctly demonstrated for the the square wave but the same form results suitable for the sinusoidal driving, with the proper fit parameters.

The observations and insights gained from the application of these periodic end-pulled forces on polymers can be generalized and applied in different experimental setups, typically associated to force spectroscopy experiments, namely optical and magnetic tweezers or atomic force microscopes. In fact, these devices are in principle able to apply either the longitudinal or the transversal drivings, as in the model here studied. 

A possible generalization of this work would be the introduction of chains pulled at a given velocity, which is typical way of performing this kind of experiments, always applying the longitudinal or transversal character of the driving. This would represent a more common approach regarding force spectroscopy experiments, whose goal consist in measuring the pull force along the recorded chain extension.

\section{Acknowledgements.}
The authors acknowledge the Grant PID2020-113582GB-I00 funded by MCIN/AEI/ 10.13039/501100011033, and the support of the Aragon Government to the Recognized group `E36\_23R F\'isica Estad\'istica y no-lineal (FENOL)'. AF also acknowledges the funds of the European Union-NextGenerationEU, and the Spanish Ministerio de Universidades through the grant BOA 139 (31185) 01/07/2021. 

\section{Appendix: Determination of the analytic MTT curve at low frequencies}
\label{appendix}
\subsection{Transversal driving.}\label{appendix_trans}
\begin{figure}[]
 \centering
 \includegraphics[width=\linewidth]{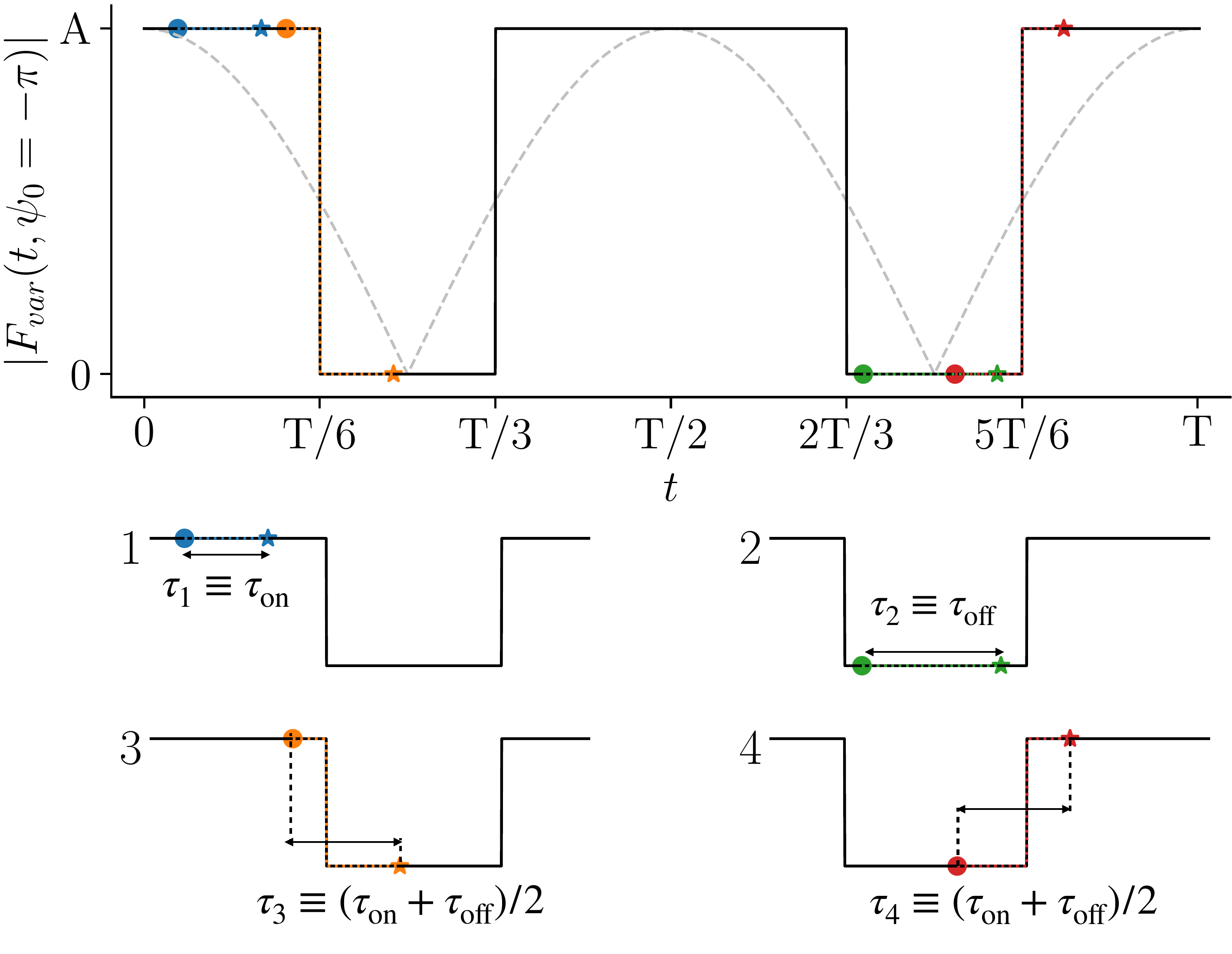}
 \caption{Schematic depiction of the evolution of the absolute value of the force with the period, both for the sinusoidal (grey) and square wave (black). Subfigures 1 to 4 depict the 4 different possible scenarios for the translocation with the square wave.}
\label{analytical_trans_depict}
\end{figure}

We consider a square wave analogous to the sinusoidal driving we apply on our system, as shown in Fig.~\ref{analytical_trans_depict}. Given the fact that for the transversal translocation both extreme values $\pm A$ aid the translocation in the same way, we have only two timescales to consider: $\tau_{\rm on}$, the time to translocate with maximum driving, and $\tau_{\rm off}$, when the force is disabled.

Depending on the value of the initial phase for each of the trajectories, 4 possible scenarios can take place. Each of them is shown in the subfigures 1, 2, 3 and 4 of Fig.~\ref{analytical_trans_depict}.

\begin{enumerate}
    \item The force starts at either $-A$ or $+A$ with a certain probability $P_A$, and remains there throughout the translocation process. In this case the TT is $\tau_{\rm on}$, and the probability to remain in the $\pm A$ value depends on the period $T$ and on the very $\tau_{\rm on}$ as $P_{A\rightarrow A}= 1-\tau_{\rm on}/(T/3)$. 
    \item The force starts at 0 with a probability $P_0$, and remains there throughout the translocation process with probability $P_{0\rightarrow 0}= 1-\tau_{\rm off}/(T/6)$.
    \item The force starts at $\pm A$, changes to 0, and remains there for the rest of the experiment. The probability switch in this case is the complementary of the case (1) that means $P_{A\rightarrow 0}= \tau_{\rm on}/(T/3)$
    \item The force starts at 0, changes to either $\pm A$, and remains there. The probability switch is then  $P_{0\rightarrow A}= \tau_{\rm off}(T/6)$, that is to say is the complementary of the case (2).
\end{enumerate}

Under these approximations, that is to say only either 0 or 1 switches in the actuating force, it is possible to write down an analytical estimation of the translocation time that takes into account the possible dynamical combinations above described, with their probability of occurrence, obtaining:
\bea 
    \langle TT \rangle &=& P_{A} \tau_{\rm on} P_{A\rightarrow A} + P_{0} \tau_{\text{off}} P_{0\rightarrow 0} + \\   &+& P_{A} \left(\frac{\tau_{\rm on}}{2} + \frac{\tau_{\rm off}}{2}\right) P_{A\rightarrow 0} + P_{0} \left(\frac{\tau_{\rm on}}{2} + \frac{\tau_{\rm off}}{2}\right) P_{0\rightarrow A}   \nonumber 
\eea
where the translocation time with one switch is considered, in the ensemble average, equal to the average of the time spent in both conditions: \emph{i.e.} $(\tau_{\rm on}/2 + \tau_{\text{off}}/2)$. Given the distribution of the phases, we find that $P_{A} = 2P_{0} = 2/3$, and the expression that results from substituting all the probabilities is
\bea
    \langle TT \rangle &=& \frac{2}{3} \tau_{\rm on} \left(1 - \frac{\tau_{\rm on}}{T/3}\right) + \frac{1}{3} \tau_{\rm off} \left(1 - \frac{\tau_{\rm off}}{T/6}\right) +    \nonumber  \\ 
    &+& \frac{2}{3} \frac{\tau_{\rm on}}{T/3} \left(\frac{\tau_{\rm on}}{2} + \frac{\tau_{\rm off}}{2}\right) + \frac{1}{3} \frac{\tau_{\rm off}}{T/6} \left(\frac{\tau_{\rm on}}{2} + \frac{\tau_{\rm off}}{2}\right) \nonumber 
    \label{low_freq_analytical}
\eea
which works very well for low frequencies, as evidenced in the curves from Fig.~\ref{scaling_transversal}.

\subsection{Longitudinal driving.}
\begin{figure}[]
 \centering
 \includegraphics[width=\linewidth]{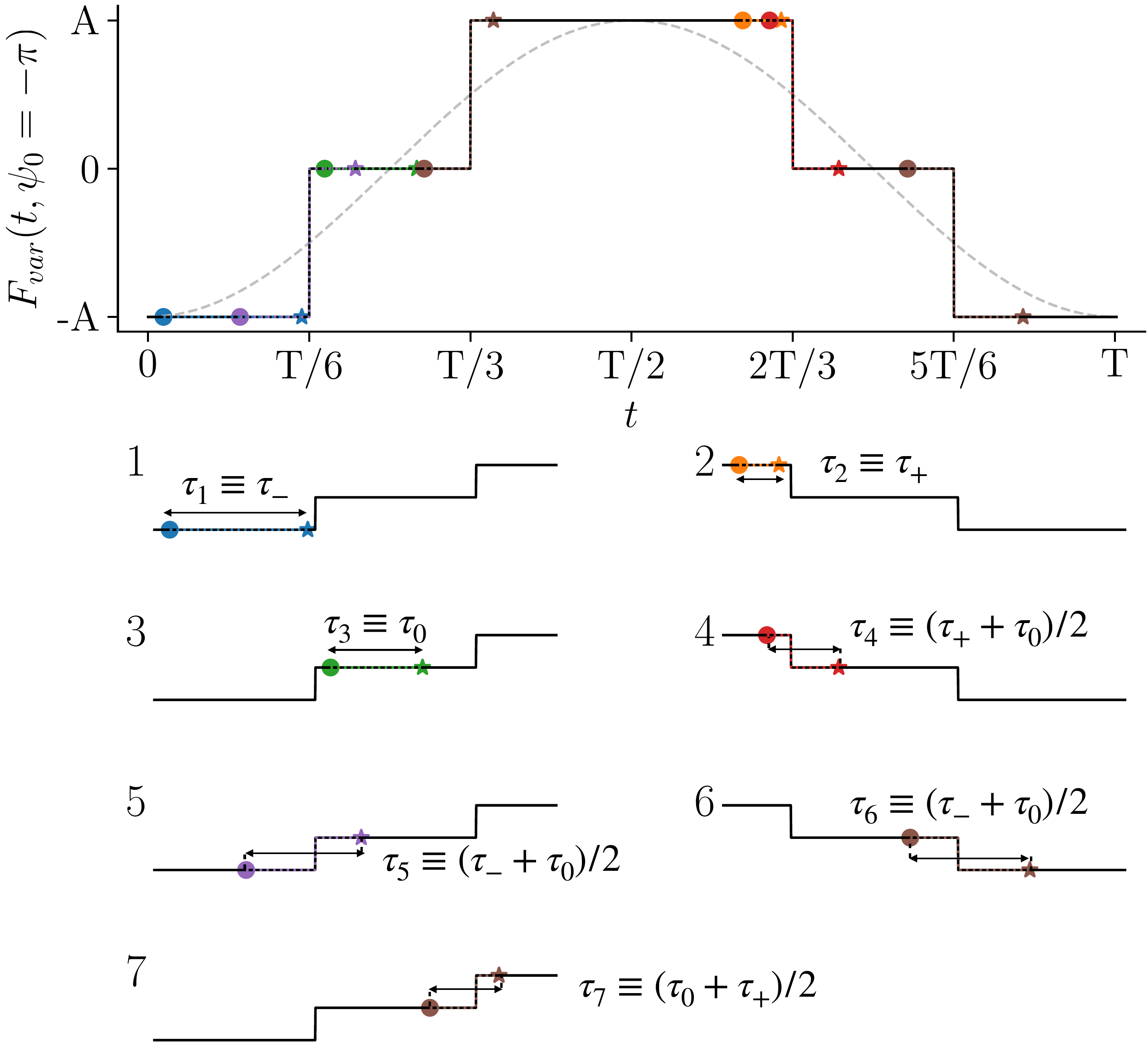}
 \caption{Schematic depiction of the evolution of the value of the force with the period, both for the sinusoidal (grey) and square wave (black). Subfigures 1 to 7 depict the 7 different possible scenarios for the translocation with the square wave.}
\label{analytical_long_depict}
\end{figure}

A square wave following the extreme values of the sinusoidal is considered for the derivation of the analytical formula. However, given that for the longitudinal translocation the values $A$ and $-A$ do not lead to the same results, three different timescales must be considered: $\tau_{\rm +}$, the time taken for the translocation to complete when the force value is $+A$, $\tau_{\rm 0}$, the time it takes when the driving is 0, and $\tau_{\rm -}$, the time necessary to translocate when the amplitude of the force is at its lowest point, $-A$. Depending on the initial phase of each trajectory, we now have six possible scenarios to take into account. All of them are contained in the different subfigures of Fig. \ref{analytical_long_depict}

\begin{enumerate}
    \item The force starts at its minimum value $-A$ with probability $P_{-A}$, and remains there throughout the translocation process. The TT is therefore $\tau_{-}$, and the probability to remain on that force value depends on the relation between that timescale and the period of the force, finding $P_{A^-\rightarrow A^-} = 1 - \tau_{-} / (T/3)$.
    \item The force starts at its maximum value $+A$ with probability $P_{(A^+)}$, and remains there throughout the translocation process. The TT is $\tau_{+}$, and the probability to remain $P_{A^+\rightarrow A^+} = 1 - \tau_{+} / (T/3)$.
    \item The force starts at 0 value value with probability $P_{0}$, and remains there throughout the translocation. The TT is $\tau_{0}$, and the probability to remain $P_{0\rightarrow 0}= 1-\tau_{0}/(T/6)$.
    \item The force starts at $+ A$, changes to 0, and remains there for the rest of the experiment. The probability to switch in this case is the complementary of the case (1), and thus $P_{A^+\rightarrow 0}= \tau_{+}/(T/3)$
    \item The force starts at $- A$, changes to 0, and remains there for the rest of the experiment. The probability to switch in this case is the complementary of (2), and thus $P_{A^-\rightarrow 0}= \tau_{0}/(T/3)$
    \item The force starts at $0$, changes to $-A$, and remains there for the rest of the experiment. Given that the phases are distributed evenly, and because of the form of the square wave, the probability switching of going to the maximum driving will be $ P_{0\rightarrow A^-}~=~ (1/2)\tau_{0} / (T/6)$.
    \item The force starts at $0$, and changes to $+A$, and remains there for the rest of the experiment. The switching probability to reach the minimum driving is the same as to go to the maximum for the symmetry reasons, {\it i.e.} is the same as point 6.: $P_{0\rightarrow A^+}~=~P_{0\rightarrow A^-}~=~ (1/2)\tau_{0} / (T/6)$.
\end{enumerate}

With these considerations, and for low enough frequencies to assume that just either zero or one changes takes place in the force, it is possible to derive an analytical expression for the average translocation time. Considering the conditional probability to start at any of those regimes and switch 0 or once into another, times their associated translocation times, we can write:
\bea
    \langle TT \rangle &=& P_{A^+} \tau_{+} P_{A^+\rightarrow A^+} + P_{0} \tau_{0} P_{0\rightarrow 0} + \nonumber \\  &+& P_{A^-} \tau_{-} P_{A^-\rightarrow A^-} +\\ \nonumber
    &+& P_{A^+} \left(\frac{\tau_{+}}{2} + \frac{\tau_{0}}{2}\right) P_{A^+\rightarrow 0} + \\ \nonumber
    &+& P_{A^-} \left(\frac{\tau_{-}}{2} + \frac{\tau_{0}}{2}\right) P_{A^-\rightarrow 0} + \\ \nonumber
    &+& P_{0} \left(\frac{\tau_{+}}{2} + \frac{\tau_{0}}{2}\right) P_{0\rightarrow A^+} + \\ \nonumber
    &+& P_{0} \left(\frac{\tau_{-}}{2} + \frac{\tau_{0}}{2}\right) P_{0\rightarrow A^-}. \nonumber 
    \label{analytical_long_1}
\eea

The translocation time associated to a switch is the averages of the two translocation timescales involved, given that what we are computing is the ensemble average over the initial phase. Substituting the values of the probabilities of jumping, and taking the probability to start in one specific state as $P_{0} = P_{A^+} = P_{A^{-}} = 1/3 $ from the distribution of the initial phases, we have the formula shown in the results section
\bea 
    \langle TT \rangle = \frac{1}{3} \tau_{+} \left(1 - \frac{\tau_{+}}{T/3}\right) + \frac{1}{3} \tau_{-} \left(1 - \frac{\tau_{-}}{T/3}\right)\nonumber\\ + \frac{1}{3} \tau_{0} \left(1 - \frac{\tau_{0}}{T/6}\right) + \frac{1}{3} \frac{\tau_{+}}{T/3} \left(\frac{\tau_{+}}{2} + \frac{\tau_{0}}{2}\right)\nonumber\\ 
    + \frac{1}{3} \frac{\tau_{-}}{T/3} \left(\frac{\tau_{-}}{2} + \frac{\tau_{0}}{2}\right) + \frac{1}{3} \frac{\tau_{0}}{T/6} \left(\frac{\tau_{0}}{2} + \frac{\tau_{+}}{4} + \frac{\tau_{-}}{4}\right).
    \label{analytical_long_3}
\eea


\begin{thebibliography}{99}
\bibitem{1996_kasianowicz_PNAS_characterization_nucleotides}Kasianowicz, J., Brandin, E., Branton, D. \& Deamer, D. Characterization of individual polynucleotide molecules using a membrane channel. {\em Proceedings Of The National Academy Of Sciences} \textbf{93}, 13770-13773 (1996)

\bibitem{2001_Ishikawa_PNAS_clpap}Ishikawa, T., Beuron, F., Kessel, M., Wickner, S., Maurizi, M. \& Alasdair C. Steven Translocation pathway of protein substrates in ClpAP protease. {\em Proceedings Of The National Academy Of Sciences} \textbf{98}, 4328-4333 (2001)

\bibitem{2007_Lebedev_EMBO_virus_translocation}Lebedev, A., Krause, M., Isidro, A., Vagin, A., Orlova, E., Turner, J., Dodson, E., Tavares, P. \& Antson, A. Structural framework for DNA translocation via the viral portal protein. {\em The EMBO Journal} \textbf{26}, 1984-1994 (2007)

\bibitem{2012_Schneider_nature_sequencing}Schneider, G. \& Dekker, C. DNA sequencing with nanopores. {\em Nature Biotechnology} \textbf{30}, 326-328 (2012)

\bibitem{2018_Lee_AV_nanopore_review}Lee, K., Park, K., Kim, H., Yu, J., Chae, H., Kim, H. \& Kim, K. Recent Progress in Solid-State Nanopores. {\em Advanced Materials} \textbf{30}, 1704680 (2018)

\bibitem{2013_Stoloff_COB_nanopore_review}Stoloff, D. \& Wanunu, M. Recent trends in nanopores for biotechnology. {\em Current Opinion In Biotechnology} \textbf{24}, 699-704 (2013)

\bibitem{2019_Mohammadreza_constant_force}Hamidabad, M. \& Abdolvahab, R. Translocation through a narrow pore under a pulling force. {\em Scientific Reports} \textbf{9}, 17885 (2019)

\bibitem{2023_Lu_review_force_current}Lu, L., Wang, Z., Shi, A., Lu, Y. \& An, L. Polymer Translocation. {\em Chinese Journal Of Polymer Science} \textbf{41}, 683–698 (2023)

\bibitem{2010_Schneider_passive_pore}Schneider, G., Kowalczyk, S., Calado, V., Pandraud, G., Zandbergen, H., Vandersypen, L. \& Dekker, C. DNA Translocation through Graphene Nanopores. {\em Nano Letters} \textbf{10}, 3163-3167 (2010)

\bibitem{2018_Hsiao_electric_field}Hsiao, P. Translocation of Charged Polymers through a Nanopore in Monovalent and Divalent Salt Solutions: A Scaling Study Exploring over the Entire Driving Force Regimes. {\em Polymers} \textbf{10}, 1229 (2018)

\bibitem{2010_Fiasconaro_Spagnolo_PB_resonant_activation}Pizzolato, N., Fiasconaro, A., Adorno, D. \& Spagnolo, B. Resonant activation in polymer translocation: new insights into the escape dynamics of molecules driven by an oscillating field. {\em Physical Biology} \textbf{7}, 034001 (2010)

\bibitem{2013_Fiasconaro_Falo_PRE_phi29}Perez-Carrasco, R., Fiasconaro, A., Falo, F. \& Sancho, J. Modeling the mechanochemistry of the $\ensuremath{\phi}29$ DNA translocation motor. {\em Phys. Rev. E} \textbf{87}, 032721 (2013)

\bibitem{2015_fiasconaro_falo_polymer_3D}Fiasconaro, A., Mazo, J. \& Falo, F. Active polymer translocation in the three-dimensional domain. {\em Physical Review E} \textbf{91}, 022113 (2015)

\bibitem{2010_Fiasconaro_Falo_PRE_mid_point}Fiasconaro, A., Mazo, J. \& Falo, F. Translocation time of periodically forced polymer chains. {\em Phys. Rev. E} \textbf{82}, 031803 (2010)

\bibitem{2011_cohen_flickering_pores}Cohen, J., Chaudhuri, A. \& Golestanian, R. Active Polymer Translocation through Flickering Pores. {\em Phys. Rev. Lett.} \textbf{107}, 238102 (2011)

\bibitem{2015_Sarabadani_time_depdendent_translocation}Sarabadani, J., Ikonen, T. \& Ala-Nissila, T. Theory of polymer translocation through a flickering nanopore under an alternating driving force. {\em The Journal Of Chemical Physics} \textbf{143}, 074905 (2015)

\bibitem{2011_Rowghanian_translocation_constant_force_pore}Rowghanian, P. \& Grosberg, A. Force-Driven Polymer Translocation through a Nanopore: An Old Problem Revisited. {\em The Journal Of Physical Chemistry B} \textbf{115}, 14127-14135 (2011)

\bibitem{2008_Neuman_force_spectroscopy}Neuman, K. \& Nagy, A. Single-molecule force spectroscopy: Optical tweezers, magnetic tweezers and atomic force microscopy. {\em Nature Methods} \textbf{5}, 491-505 (2008)

\bibitem{2022_Paun_polymer_constant_force}Paun, M., Paun, V. \& Paun, V. Polymer Translocation through Nanometer Pores. {\em Polymers} \textbf{14}, 1166 (2022)

\bibitem{2009_Sun_point_pore}Sun, L., Cao, W. \& Luo, M. Free energy landscape for the translocation of polymer through an interacting pore. {\em The Journal Of Chemical Physics} \textbf{131}, 194904 (2009)

\bibitem{2011_Fiasconaro_Falo_JSM_end_pulled_dichotomic}Fiasconaro, A., Mazo, J. \& Falo, F. Translocation of a polymer chain driven by a dichotomous noise. {\em Journal Of Statistical Mechanics: Theory And Experiment} \textbf{2011}, 11002 (2011)

\bibitem{2012_Fiasconaro_Falo_MM_ATP}Fiasconaro, A., Mazo, J. \& Falo, F. Michaelis–Menten dynamics of a polymer chain out of a dichotomous ATP-based motor. {\em New Journal Of Physics} \textbf{14}, 023004 (2012)

\bibitem{2017_Fiasconaro_SR_ATP_motor}Fiasconaro, A., Mazo, J. \& Falo, F. Active translocation of a semiflexible polymer assisted by an ATP-based molecular motor. {\em Scientific Reports} \textbf{7}, 4188 (2017)

\bibitem{2013_Bhattacharya_driven_pore_force_2D}Adhikari, R. \& Bhattacharya, A. Driven translocation of a semi-flexible chain through a nanopore: A Brownian dynamics simulation study in two dimensions. {\em The Journal Of Chemical Physics} \textbf{138}, 204909 (2013)

\bibitem{2012_Saito_pore_force}Saito, T. \& Sakaue, T. Process time distribution of driven polymer transport. {\em Phys. Rev. E} \textbf{85}, 061803 (2012)

\bibitem{2018_Sarabadani_end_pulled}Sarabadani, J. \& Ala-Nissila, T. Theory of pore-driven and end-pulled polymer translocation dynamics through a nanopore: an overview. {\em Journal Of Physics: Condensed Matter} \textbf{30}, 274002 (2018)

\bibitem{2006_Ritort_SME_translocation}Ritort, F. Single-molecule experiments in biological physics: methods and applications. {\em Journal Of Physics: Condensed Matter} \textbf{18}, R531 (2006)

\bibitem{2006_Keyser_solid_state_nanopore}Keyser, U., Koeleman, B., Dorp, S., Krapf, D., Smeets, R., Lemay, S., Dekker, N. \& Dekker, C. Direct force measurements on DNA in a solid-state nanopore. {\em Nature Physics} \textbf{2} 473–477 (2006)

\bibitem{1996_Sung_free_energy_analytical}Sung, W. \& Park, P. Polymer Translocation through a Pore in a Membrane. {\em Phys. Rev. Lett.}. \textbf{77}, 783-786 (1996)

\bibitem{2011_Saito_analytical_scaling}Saito, T. \& Sakaue, T. Dynamical diagram and scaling in polymer driven translocation. {\em The European Physical Journal E} \textbf{34}, 1-8 (2011)

\bibitem{2012_dubbeldam_forced_scaling}Dubbeldam, J., Rostiashvili, V., Milchev, A. \& Vilgis, T. Forced translocation of a polymer: Dynamical scaling versus molecular dynamics simulation. {\em Phys. Rev. E} \textbf{85}, 041801 (2012)

\bibitem{2009_Sun_Free_energy}Sun, L., Cao, W. \& Luo, M. Free energy landscape for the translocation of polymer through an interacting pore. {\em The Journal Of Chemical Physics} \textbf{131}, 194904 (2009)

\bibitem{2016_muthukumar_polymer}Muthukumar, M. Polymer translocation. (CRC press, 2016)

\bibitem{1999_muthukumar_polymer_translocation}Muthukumar, M. Polymer translocation through a hole. {\em The Journal Of Chemical Physics} \textbf{111}, 10371 (1999)

\bibitem{2018_Menais_PRE_end_pulled}Menais, T. Polymer translocation under a pulling force: Scaling arguments and threshold forces. {\em Phys. Rev. E} \textbf{97}, 022501 (2018)

\bibitem{2016_Sakaue_escalado_pubmed}Sakaue, T. Dynamics of Polymer Translocation: A Short Review with an Introduction of Weakly-Driven Regime. {\em Polymers} \textbf{8}, 424 (2016)

\bibitem{2022_Sarabadani_tension_propagation}Sarabadani, J., Metzler, R. \& Ala-Nissila, T. Driven polymer translocation into a channel: Isoflux tension propagation theory and Langevin dynamics simulations. {\em Phys. Rev. Res.} \textbf{4}, 033003 (2022)

\bibitem{2018_fiasconaro_falo_force_spectroscopy}Fiasconaro, A. \& Falo, F. Force spectroscopy analysis in polymer translocation. {\em Physical Review E} \textbf{98}, 062501 (2018)

\bibitem{2022_Fiasconaro_Falo_Polymer_pore_explicit}Fiasconaro, A., Díez-Señorans, G. \& Falo, F. End-pulled polymer translocation through a many-body flexible pore. {\em Polymer}. \textbf{259}, 125305 (2022)

\bibitem{2023_singh_chauhan_sequencing_review}Singh, S., Chauhan, K., Bharadwaj, A., Kishore, V., Laux, P., Luch, A. \& Singh, A. Polymer Translocation and Nanopore Sequencing: A Review of Advances and Challenges. {\em International Journal Of Molecular Sciences} \textbf{24}, 6153 (2023)






\bibitem{2012_Ikonen_RA}Ikonen, T., Shin, J., Sung, W. \& Ala-Nissila, T. Polymer translocation under time-dependent driving forces: Resonant activation induced by attractive polymer-pore interactions. {\em The Journal Of Chemical Physics} \textbf{136}, 205104 (2012)

\bibitem{2014_suhonen_PRE_polymer_translocation}Suhonen, P., Kaski, K. \& Linna, R. Criteria for minimal model of driven polymer translocation. {\em Phys. Rev. E} \textbf{90}, 042702 (2014)

\bibitem{1992_Doering_PRE_Resonant_activation}Doering, C. \& Gadoua, J. Resonant activation over a fluctuating barrier. {\em Phys. Rev. Lett.} \textbf{69}, 2318-2321 (1992)

\bibitem{1996_doi_introduction}Doi, M. Introduction to polymer physics. (Oxford university press,1996)

\bibitem{2013_Ikonen}Ikonen, T., Bhattacharya, A., Ala-Nissila, T. \& Sung, W. Influence of pore friction on the universal aspects of driven polymer translocation. {\em Europhysics Letters} \textbf{103}, 38001 (2013)


\bibitem{2011_Fiasconaro_Spagnolo_RA}Fiasconaro, A. \& Spagnolo, B. Resonant activation in piecewise linear asymmetric potentials. {\em Phys. Rev. E} \textbf{83}, 041122 (2011)


\end{thebibliography}
\end{document}